\documentclass[a4paper,11pt]{article}
\pdfoutput=1 

\usepackage{amsmath} 
\allowdisplaybreaks
\usepackage{jheppub} 
\usepackage{empheq}
\usepackage{bm}
\usepackage{slashed}
\usepackage[T1]{fontenc} 
\usepackage[all]{xy}
\usepackage{rotating}
\usepackage{mathtools}
\usepackage{bbm}
\usepackage{dcolumn}
\usepackage{multirow}
\usepackage{mathrsfs}
\usepackage{arydshln}
\usepackage{verbatim}
\usepackage{graphics}
\usepackage{longtable}
\usepackage{dsfont}
\usepackage{physics}
\usepackage{kotex}
\usepackage{longtable}
\usepackage{makecell}
\usepackage{float}
\usepackage{comment}

\title{\boldmath Non-hyperbolic 3-manifolds and bulk field theories for supersymmetric/$W_N$ minimal models}

\author[a,b]{Seungjoo Baek,}
\author[a]{Heesu Kang}

\affiliation[a]{
	Department of Physics and Astronomy $\&$ Center for Theoretical Physics,
	\\
	Seoul National University, 1 Gwanak-ro, Seoul 08826, Korea}
\affiliation[b]{Department of Physics, Princeton University, Princeton, NJ 08544, USA}

\emailAdd{sb2880@princeton.edu}
\emailAdd{heesu0434@snu.ac.kr}
\abstract{Building on the work of Gang, Kang, and Kim \cite{Gang:2024}, we propose 3D bulk dual field theories for 2D $\mathcal{N}=1$ supersymmetric minimal models $SM(P, Q)$ and $W_{N}$ algebra minimal models $W_{N}(P, Q)$. We associate to $SM(P, Q)$ a Seifert fibered space $S^2((P,P-R),(Q,S),(3,1))$ with $PS-QR=2$, and for $W_{N}(P, Q)$ a Seifert fibered space $S^2((P,P{-}R),(Q,S),(N{+}1,-2N{-}1))$ with $PS-QR=1$, and realize the bulk theory via the 3D-3D correspondence. For the unitary series, the bulk theory flows in the IR to a gapped phase which, under suitable boundary conditions, supports the unitary chiral minimal model on the boundary. For the non-unitary series, the bulk theory flows to the 3D $\mathcal{N}=4$ superconformal field theory whose topological twist yields a non-unitary topological field theory supporting the non-unitary chiral minimal model on the boundary under appropriate boundary conditions. We also propose UV gauge theory descriptions of the bulk theories obtained by gluing $T[SU(n)]$ building blocks. For $SM(P, Q)$, we provide non-trivial consistency checks---matching between various bulk partition functions and boundary conformal data---while for  $W_N(P, Q)$, we present preliminary checks and leave further consistency checks for future work.}

\begin{document} 
\maketitle
\flushbottom	

\section{Introduction}

Two-dimensional conformal field theories (CFTs) play a central role across various physical phenomena, ranging from second-order critical phenomena and the string worldsheet theories to RG fixed points of 2D QFTs. Within this landscape, rational CFTs (RCFTs)---distinguished by a finite number of irreducible representations---possess a mathematical framework called a modular tensor category, which can be realized by a 3D topological quantum field theory (TQFT) through the bulk-boundary correspondence. An archetypal example is the Chern-Simons/Wess-Zumino-Witten (CS/WZW) correspondence, which identifies the Hilbert space of the Chern-Simons theory on a 3-manifold with boundary with the conformal blocks of the WZW theory on that boundary \cite{Witten:1988hf}. Nevertheless, a general bulk field-theoretic description for RCFTs beyond the CS/WZW correspondence---especially for non-unitary RCFTs---remains poorly understood. The bulk theory construction for non-unitary RCFTs has been studied only recently \cite{Gang:2018huc, Gang:2023rei, Baek:2025, Dedushenko:2020, Gang:2021hrd, Gang:2023, Ferrari:2023, Gang:2023ano, Dedushenko:2023, Gang:2024}.
    
In this paper, we extend the work of \cite{Gang:2024} to study 3D bulk field theory descriptions of the 2D $\mathcal{N}=1$ supersymmetric minimal models $SM(P, Q)$ and the 2D $W_N$ algebra minimal models $W_{N}(P, Q)$.  The bulk field theory is constructed as follows. Using 3D-3D correspondence \cite{Terashima:2011qi,Dimofte:2011ju,Gang:2018wek}, we build a 3D class R theory (or its bosonization) for the Seifert fibered space $S^2(\vec{p},\,\vec{q})$. For the unitary minimal model, bulk theory flows to the gapped phase in the IR, and supports a unitary rational chiral algebra under an appropriate boundary condition. For the non-unitary minimal model, bulk theory flows to the 3D $\mathcal{N}=4$ SCFT in the IR. After topological twisting, the theory supports a non-unitary rational chiral algebra under an appropriate boundary condition. This procedure can be summarized as follows:
\begin{equation}
    \begin{aligned}
        &\text{Seifert fibered manifolds}\xrightarrow{\text{3D-3D}}\text{gapped theories or 3D }\mathcal{N}=4\text{ rank-0 SCFTs}\\
        &\xrightarrow{\text{bulk-boundary}}\text{2D unitary(or non-unitary) }\mathcal{N}=1\text{ minimal models, }W_{N}\text{ minimal models}
    \end{aligned}
\end{equation}
We check the proposed correspondence using the known dictionaries. For example, for $SM(P, Q)$, we compare the topological invariants of the bulk theory with the modular data of the boundary CFT. But for $W_{N}(P, Q)$, we manage to compute only the partial bulk data, so our proposal is more speculative. We leave further calculations for $W_{N}(P, Q)$ to future work.

The rest of this paper is organized as follows. In Section~\ref{sec:sm}, we first review the basics of $\mathcal{N}=1$ supersymmetric minimal model $SM(P, Q)$ and dictionary for the non-unitary bulk-boundary correspondence. We then propose the bulk field theory of $SM(P, Q)$ and its gauge theory description with $T[SU(2)]$ building blocks. We check the correspondence using the dictionary in Tables~\ref{tab:ferdict} and \ref{tab:bosdict}. In Section~\ref{sec:sm}, we review the basics of $W_{N}$ algebra minimal model $W_{N}(P, Q)$ and propose the bulk field theory. We then present some supporting calculations. In the appendix, we provide details for the computation of the relevant partition functions.

\section{Bulk field theory for 2D $\mathcal{N}=1$ minimal model $SM(P, Q)$} \label{sec:sm}

\subsection{2D $\mathcal{N}=1$ minimal model $SM(P, Q)$}
$\mathcal{N}=1$ supersymmetric minimal models $SM(P, Q)(=SM(Q,P))$ are labeled by two integers $P$ and $Q$ satisfying
\begin{equation}
    2\leq Q<P\,,\,P-Q\in 2\mathbb{Z}\text{ and }\gcd\left(P,\frac{P-Q}{2}\right)=1\,.
\end{equation}
$SM(P, Q)$ is unitary if and only if $P=Q+2$. Conformal primaries are labeled as $\mathcal{O}_{(s,t)}$ with $1\leq s\leq Q-1$ and $1\leq t\leq P-1$. $\mathcal{O}_{(1,1)}$ is the identity operator and there is an equivalence relation $\mathcal{O}_{(s,t)}=\mathcal{O}_{(Q-s,P-t)}$. $s+t$ even and odd corresponds to the NS(Neveu-Schwarz, antiperiodic boundary condition) sector and R(Ramond, periodic boundary condition) sector, respectively. Central charge and conformal dimensions of $SM(P, Q)$ are
\begin{equation}
\label{eq:smtmat}
    \begin{aligned}
        &c=\frac{3}{2}\left(1-\frac{2(P-Q)^{2}}{PQ}\right),\\
        &h_{(s,t)}=\frac{(Ps-Qt)^{2}-(P-Q)^{2}}{8PQ}+\frac{2\epsilon_{s-t}-1}{16}\,,\quad\epsilon_{a}=
        \begin{cases}\frac{1}{2}& a\in 2\mathbb{Z}\text{ (NS sector)}\\
        1& a\in 2\mathbb{Z}+1\text{ (R sector)}
    \end{cases}\,.
    \end{aligned}
\end{equation}
The conformal characters of the NS sector primaries are
\begin{equation}
\label{eq:smchar}
    \chi_{(s,t)}=q^{h_{(s,t)}-\frac{c}{24}}\frac{(-q^{\frac{1}{2}};q)_{\infty}}{(q;q)_{\infty}}\sum_{l\in\mathbb{Z}}\left(q^{\frac{l(lPQ+sP-tQ)}{2}}-q^{\frac{(lQ+s)(lP+t)}{2}}\right)\,,
\end{equation}
where we introduced the $q$-Pochhammer symbol 
\begin{equation}
    (a;q)_{n}\coloneq\prod_{l=0}^{n-1}(1-aq^{l})\,,\,(q)_{n}\coloneq(q;q)_{n}=\prod_{l=1}^{n}(1-q^{l})\,.
\end{equation}
Note that on the torus, the (NS, NS) sector maps to itself under $S$ and $T^{2}$ transformations. In particular, under the modular $S$ transformation, NS characters transform as follows:
\begin{equation}
    \chi_{\alpha}(\tilde{q})=\sum_{\beta}S^{\text{NS}}_{\alpha\beta}\chi_{\beta}(q)\quad\text{where }q\coloneq e^{2\pi i\tau}\text{ and }\tilde{q}\coloneq e^{2\pi i(-1/\tau)}\,,
\end{equation}
with $\alpha$ and $\beta$ labeling the NS primaries. The modular S matrix of the (NS, NS) sector is
\begin{equation}
\label{eq:smsmat}
    S^{\text{NS}}_{(s_{1},t_{1})(s_{2},t_{2})}=\frac{2}{\sqrt{PQ}}\left(\cos{\frac{2\pi\lambda_{1}\lambda_{2}}{4PQ}}-\cos{\frac{2\pi\bar{\lambda}_{1}\lambda_{2}}{4PQ}}\right)\,,
\end{equation}
with $\lambda_{i}=Qt_{i}-Ps_{i},\,\bar{\lambda}_{i}=Qt_{i}+Ps_{i}$.

\subsection{Bulk field theory as $T_{\text{irred}}[M]$ with $M=S^2((P,P-R),(Q,S),(3,1))$}\label{sec:smbulkdual}
Recently, it was found in \cite{Gang:2021hrd} that non-unitary RCFTs live on the boundary of topologically twisted 3d $\mathcal{N}=4$ rank-0 theory. Here, rank-0 means that the theory has an empty Coulomb and Higgs branch. The boundary RCFT generally depends on the holomorphic boundary condition $\mathbb{B}$.
We define an RCFT $R[\mathcal{T},\mathds{B}]$ as follows. When $R[\mathcal{T},\mathds{B}]$ is unitary, it represents a theory living on the boundary of $\mathcal{T}$ with the boundary condition $\mathds{B}$. For the non-unitary case, $R[\mathcal{T},\mathds{B}]$ is living on the boundary (with boundary condition $\mathds{B}$) of the non-unitary TQFT $\mathcal{T}^{\text{top}}$, which is a topological twist of a 3D $\mathcal{N}=4$ rank-0 SCFT $\mathcal{T}$. From this definition, $\mathcal{T}_{SM(P, Q)}$ is defined as a bulk theory such that $R[\mathcal{T}_{SM(P, Q)},\mathds{B}]=SM(P, Q)$ for some proper $\mathds{B}$.
Also, we will denote the bulk field theory of the bosonized $SM(P, Q)$ as $\tilde{\mathcal{T}}_{SM(P, Q)}$. 
\begin{equation}
    \begin{aligned}
        &\textbf{Def: 3D }\mathcal{T}_{SM(P, Q)},\,\widetilde{\mathcal{T}}_{SM(P, Q)}\textbf{ theories are defined as}\\
        &\text{For }|P-Q|=2,\,\mathcal{T}_{SM(P, Q)}(\text{resp. }\widetilde{\mathcal{T}}_{SM(P, Q)})\text{ is a 3D unitary spin-TQFT(resp. TQFT) with}\\
        &\mathcal{T}_{SM(P, Q)}\xrightarrow{\text{at boundary with a proper }\mathbb{B}}SM(P, Q),\\
        &\widetilde{\mathcal{T}}_{SM(P, Q)}\xrightarrow{\text{at boundary with a proper }\mathbb{B}}\text{bosonized }SM(P, Q)\,.\\
        &\text{For }|P-Q|>2,\,\mathcal{T}_{SM(P, Q)},\,\widetilde{\mathcal{T}}_{SM(P, Q)}\text{ is a }\mathcal{N}=4\text{ 3D rank-0 SCFT with}\\
        &\mathcal{T}_{SM(P, Q)}\xrightarrow{\text{top'l twist}}\text{non-unitary TQFT }\mathcal{T}^{\text{top}}_{SM(P, Q)}\xrightarrow{\text{at bdy. with a proper }\mathbb{B}}SM(P, Q),\\
        &\widetilde{\mathcal{T}}_{SM(P, Q)}\xrightarrow{\text{top'l twist}}\text{non-unitary TQFT }\widetilde{\mathcal{T}}^{\text{top}}_{SM(P, Q)}\xrightarrow{\text{at bdy. with a proper }\mathbb{B}}\text{bosonized }SM(P, Q)\,.
    \end{aligned}
\end{equation}

Basic dictionaries of the bulk-boundary correspondence are given in the first and second columns of Table~\ref{tab:ferdict} and \ref{tab:bosdict}. For more details see \cite{Gang:2021hrd,Gang:2023rei,Cho:2020ljj}. We will realize the bulk theory as an IR fixed point of 3D $\mathcal{N}\geq 2$ gauge theory. IR dynamics can be probed using BPS partition functions, which are RG-invariant. 3d $\mathcal{N}=4$ theories have $SU(2)^{C}\times SU(2)^{H}$ R-symmetry. We will denote their Cartans as $J_{3}^{C},\,J_{3}^{H}\in\mathbb{Z}/2$, respectively. For UV localization computation, we will use the $\mathcal{N}=2$ subalgebra manifest in UV, whose R-symmetry is generated by $J_{3}^{C}+J_{3}^{H}$. Then $J_{3}^{C}-J_{3}^{H}$ becomes a $U(1)_{A}$ flavor symmetry. We can mix the R-symmetry with a flavor symmetry:
\begin{equation}
    R_{\nu}=(J_{3}^{C}+J_{3}^{H})+\nu (J_{3}^{C}-J_{3}^{H})\,.
\end{equation}
Partition functions depend on the real mass $M$ associated with $U(1)_{A}$ and the mixing parameter $\nu$. $(M,\nu)=(0,0)$ corresponds to the superconformal point. For 3d $\mathcal{N}=4$ rank-0 SCFT, setting $(M,\nu)=(0,-1),\,(0,+1)$ gives the partition function of A-twisted and B-twisted theory, respectively. We will make use of various partition functions; first, consider squashed three-sphere partition function $\mathcal{Z}^{S_{b}^{3}}$ \cite{Kapustin:2009kz,Jafferis:2010un,Hama:2010av}, where
\begin{equation}
\label{eq:squashed}
    S_{b}^{3}\coloneq\{(z,w)\in\mathbb{C}^{2}\,:\,b^{2}|z|^{2}+b^{-2}|w|^{2}=1\}\,.
\end{equation}
For future use, we define
\begin{equation}
    \mathcal{Z}^{\text{con}}_{b}\coloneq\mathcal{Z}^{S_{b}^{3}}\rvert _{(M,\nu)\rightarrow(0,0)}\,.
\end{equation}
Partition functions on various 3-manifolds can be computed as follows \cite{Benini:2015noa,Benini:2016hjo,Closset:2016arn,Closset:2018ghr}:
\begin{equation}
    \mathcal{Z}^{\mathcal{M}_{g,p}}=\sum_{\vec{x}_{\alpha}\in\mathcal{S}_{\text{BE}}}\mathcal{H}(\vec{x}_{\alpha})^{g-1}\mathcal{F}(\vec{x}_{\alpha})^{p}\,.
    \label{eq:ptn}
\end{equation}
Here, $\vec{x}_{\alpha}$ are the Bethe vacua, which are ground states on a two-torus $\mathbb{T}^{2}$ when the bulk theory is a topological field theory. $\mathcal{M}_{g,p}$ is a $S^{1}$-bundle of degree $p$ over a Riemann surface $\Sigma_{g}$, and $\mathcal{H}$ and $\mathcal{F}$ are handle-gluing and fibering operators, respectively. Note that, to compute the partition function on $\mathcal{M}_{g,p}$ using localization, we need a supersymmetry-preserving background. When $p$ is even, we have two choices of background depending on the spin structure of $S^{1}$(periodic or antiperiodic boundary condition), while for odd $p$, only periodic boundary condition is allowed. For bosonic TQFT whose partition function is independent of the spin structure, we can apply \eqref{eq:ptn} for even and odd $p$. For fermionic TQFT, we can apply \eqref{eq:ptn} only for even $p$.

The bulk field theory $\mathcal{T}$ for an RCFT $R[\mathcal{T},\mathds{B}]$ can be obtained via 3D-3D correspondence. From 3D-3D correspondence, one can construct a 3D class R theory $T_{\text{irred}}[M]$ associated with a closed 3-manifold $M$. 
The theory is believed to describe an effective 3D field theory of 6D $A_{1}\,\mathcal{N}=(2,0)$ superconformal field theory compactified on the 3-manifold $M$. Also, we consider $T_{\text{irred}}[M]$, which only sees irreducible $SL(2,\mathbb{C})$(or $PSL(2,\mathbb{C})$) flat connections on $M$ \cite{Chung:2014qpa}. Note that many previous studies were on the $T_{\text{full}}[M]$, which sees all flat connections \cite{Cheng:2018,Gadde:2013,Gukov:2017,Gukov:2017kmk,Eckhard:2019jgg,Chung:2021,Assel:2022row,Chung:2024}. For the difference between $T_{\text{irred}}$ and $T_{\text{full}}$ see \cite{Gang:2018wek}. In addition, $T_{\text{irred}}[M]$ depends on the polarization choice of the 6D $\mathcal{N}=(2,0)$ theory, which can be labeled by a subgroup $H\subset H^{1}(M,\mathbb{Z}_{2})$ \cite{Gang:2018wek,Eckhard:2019jgg}. $T_{\text{irred}}[M;H]$ has $H$ as a 0-form flavor symmetry. In the following, our convention will be
\begin{equation}
    \begin{aligned}
        &T_{\text{irred}}[M]\coloneq T_{\text{irred}}[M;H=H^{1}(M,\mathbb{Z}_{2})]\,,\\
        &\widetilde{T}_{\text{irred}}[M]\coloneq T_{\text{irred}}[M;H=\emptyset]\,.
    \end{aligned}
\end{equation}
$\widetilde{T}_{\text{irred}}[M]$ has $H^{1}(M,\mathbb{Z}_{2})$ as a 1-form symmetry, and by gauging this $H^{1}(M,\mathbb{Z}_{2})$ 1-form symmetry we can obtain $T_{\text{irred}}[M]$. Also, we will denote the boundary chiral RCFTs corresponding to $T_{\text{irred}}[M]$ and $\widetilde{T}_{\text{irred}}[M]$ as $\chi\mathcal{R}[M;\mathbb{B}]$ and $\widetilde{\chi\mathcal{R}}[M;\mathbb{B}]$, respectively.  Throughout the paper, we will consider Seifert fibered spaces $M=S^{2}(\vec{p},\vec{q})$. IR phases were empirically analyzed in \cite{Choi:2022dju}:
\begin{equation}
    \begin{aligned}
        &T_{\text{irred}}\,(\text{or }\widetilde{T}_{\text{irred}})[M=S^{2}((p_{1},q_{1}),(p_{2},q_{2}),(p_{3},q_{3}))]\\
        &\xrightarrow{\text{in the IR}}
        \begin{cases}
            \text{a unitary TQFT,}& q_{i}=\pm 1\pmod{p_{i}}\quad\forall i=1,2,3,\\
            \text{a rank-0 SCFT,}&\text{otherwise.}
        \end{cases}
    \end{aligned}
\end{equation}
Combining the 3D-3D correspondence and the bulk-boundary correspondence, our correspondence can be summarized as follows:
\begin{equation}
    M=S^{2}(\vec{p},\vec{q})\xrightarrow{\text{3D-3D correspondence}}T_{\text{irred}}(\text{resp. }\widetilde{T}_{\text{irred}})[M]\xrightarrow{\text{bulk-boundary}}\chi\mathcal{R}(\text{resp. }\widetilde{\chi\mathcal{R}})[M;\mathbb{B}]\,.
\end{equation}
\begin{table}
\centering
\begin{tabular}{ |c|c|c| } 
 \hline
 2D chiral RCFT $\chi\mathcal{R}[M;\mathbb{B}]$ & $T_{\text{irred}}[M]$ & $PSL(2,\mathbb{C})$ CS on $M$ \\ 
 \hline
 \makecell{NS-sector primaries\\ $\mathcal{O}_{\alpha=0,\cdots,N-1}$} & Bethe-vacuum $\vec{x}_{\alpha}\in\mathcal{S}_{\text{BE}}$ & \makecell{ $\rho^{PSL}$ with $w_{2}(\rho^{PSL})=0$\\(multiplicity $|\text{Inv}(\rho^{PSL})|$)} \\ 
 \hline
 Conformal dimension $h_{\alpha}$&$e^{4\pi ih_{\alpha}}=\left(\frac{\mathcal{F}^{\text{top}}(\vec{x}_{\alpha})}{\mathcal{F}^{\text{top}}(\vec{x}_{\alpha=0})}\right)^{2}$&$e^{4\pi ih_{\alpha}}=e^{4\pi i(CS[\rho^{PSL}_{\alpha=0}]-CS[\rho^{PSL}_{\alpha}])}$\\
 \hline
 $\left(S^{\text{NS}}_{0\alpha}\right)^{2}$&$(\mathcal{H}^{\text{top}}(\vec{x}_{\alpha}))^{-1}$&\makecell{$(2\text{Tor}[\rho^{PSL}]|\text{Inv}(\rho^{PSL})|^{2})^{-1}$\\$\times|H^{1}(M,\mathbb{Z}_{2})|$}\\
 \hline
 $\min_{\alpha}\{|S^{\text{NS}}_{0\alpha}|\}$&$e^{-F}=|\mathcal{Z}^{\text{con}}_{b=1}|$&$\min_{\alpha}\{\sqrt{\text{(above cell)}}\}$\\
 \hline
\end{tabular}
\caption{Basic dictionaries (adopted from \cite{Gang:2024}) for the correspondence among non-hyperbolic 3-manifolds $M$, 3d bulk theories $T_{\text{irred}}[M]$, and 2D chiral RCFTs $\chi\mathcal{R}[M]$ for $M=S^{2}((\vec{p},\vec{q}))$ with $H^{1}(M,\mathbb{Z}_{2})=\mathbb{Z}_2$. $\rho^{PSL}\in\chi_{\text{irred}}^{PSL}[M]$ is an irreducible flat connection \eqref{eq:irredflat}. $w_{2}(\rho^{PSL})$ denotes the second Stiefel-Whitney class of $\rho^{PSL}$. Inv$[\rho^{PSL}]$ is defined in \eqref{eq:inv}. The superscripts `top'$=A$ (or $B$) and `con' denote the partition function in $A$ (or $B$)-twisting limit and at the superconformal point of rank-0 SCFT, respectively. $S_{\alpha\beta}$ denotes modular S-matrix. $\mathcal{H}$ and $\mathcal{F}$ are handle-gluing and fibering operators. CS$[\rho]$ and Tor$[\rho]$ denote the Chern-Simons invariant and the adjoint Reidemeister torsion of an irreducible flat connection $\rho^{PSL}$.}
\label{tab:ferdict}
\end{table}
\begin{table}
\centering
\begin{tabular}{ |c|c|c| } 
 \hline
 2D chiral RCFT $\widetilde{\chi\mathcal{R}}[M;\mathbb{B}]$ & $\widetilde{T}_{\text{irred}}[M]$ & $PSL(2,\mathbb{C})$ CS on $M$ \\ 
 \hline
 \makecell{Bosonic mother theory\\ primaries $\mathcal{O}_{\alpha=0,\cdots,\widetilde{N}-1}$} & Bethe-vacuum $\vec{x}_{\alpha}\in\mathcal{S}_{\text{BE}}$ & $\{\rho^{PSL}\otimes\eta\}$ \\ 
 \hline
 Conformal dimension $h_{\alpha}$&$e^{2\pi ih_{\alpha}}=\frac{\mathcal{F}^{\text{top}}(\vec{x}_{\alpha})}{\mathcal{F}^{\text{top}}(\vec{x}_{\alpha=0})}$&\makecell{$e^{2\pi ih_{\alpha}}=e^{2\pi i(CS[(\rho^{PSL}\otimes\eta)_{\alpha=0}]-CS[(\rho^{PSL}\otimes\eta)_{\alpha}])}$\\(for anyons with $w_{2}(\rho^{PSL})=0$)\\$e^{4\pi ih_{\alpha}}=e^{4\pi i(CS[(\rho^{PSL}\otimes\eta)_{\alpha=0}]-CS[(\rho^{PSL}\otimes\eta)_{\alpha}])}$\\(for anyons with nontrivial $w_{2}(\rho^{PSL})$)}\\
 \hline
 $\left(S_{0\alpha}\right)^{2}$&$(\mathcal{H}^{\text{top}}(\vec{x}_{\alpha}))^{-1}$&\makecell{$(2|H^{1}(M,\mathbb{Z}_{2})|\text{Tor}[\rho^{PSL}])^{-1}$}\\
 \hline
 $\min_{\alpha}\{|S_{0\alpha}|\}$&$e^{-F}=|\mathcal{Z}^{\text{con}}_{b=1}|$&$\min_{\alpha}\{\sqrt{\text{(above cell)}}\}$\\
 \hline
\end{tabular}
\caption{Basic dictionaries (adopted from \cite{Gang:2024}) for the correspondence among non-hyperbolic 3-manifolds $M$, 3d bulk theories $\widetilde{T}_{\text{irred}}[M]$, and 2D chiral RCFTs $\widetilde{\chi\mathcal{R}}[M]$ for $M=S^{2}((\vec{p},\vec{q}))$ with $H^{1}(M,\mathbb{Z}_{2})=\mathbb{Z}_2$.}
\label{tab:bosdict}
\end{table}
Basic dictionaries of the correspondence are given in table~\ref{tab:ferdict} and \ref{tab:bosdict}. Refer to \cite{Gang:2019,Gang:2019uay,Benini:2019dyp,Cho:2020ljj,Cui:2021lyi,Bonetti:2024} for details. First, note that the $PSL(2,\mathbb{C})$ flat connections on $M$ can be alternatively described by a homomorphism as below:
\begin{equation}
\label{eq:irredflat}
    \begin{aligned}
        \chi_{\text{irred}}^{PSL}[M]=\{&\rho\in\text{Hom}[\pi_{1}(M)\rightarrow PSL(2,\mathbb{C})]\,:\,\dim H(\rho)=0\}/\sim\,,\\
        &\text{where }H(\rho)\coloneq\{g\in PSL(2,\mathbb{C})\,:\,[g,\rho (a)]=0\,\,\forall a\in\pi_{1}(M)\}\,.
    \end{aligned}
\end{equation}
The equivalence relation is up to $SL(2,\mathbb{C})$ conjugation, and the condition $\dim H(\rho)=0$ reflects the irreducibility. In this paper, we won't distinguish the homomorphism $\rho$ and the corresponding flat connection $\mathcal{A}_{\rho}$. In Table~\ref{tab:ferdict} and \ref{tab:bosdict}, $\rho^{PSL}$ is an irreducible $PSL(2,\mathbb{C})$ flat connection on $M$, and $\eta\in H^{1}(M,\mathbb{Z}_{2})$ is the $\mathbb{Z}_{2}$ flat connection. In Table~\ref{tab:bosdict}, 1-form symmetry generating anyons correspond to $A_{\eta}=\rho_{\alpha=0}^{PSL}\otimes\eta$, where $\rho_{\alpha=0}^{PSL}$ is a flat connection such that $\rho_{\alpha=0}^{PSL}\otimes 1$ corresponds to the trivial anyon. Dictionaries are valid only for the 3-manifold with $H^{1}(M,\mathbb{Z}_{2})=\mathbb{Z}_2$, such that $\mathbb{Z}_2$ 1-form symmetry is fermionic in the sense that $A_{\eta}(\eta\neq 1)$ has a topological spin $\frac{1}{2}$. For the flat connection $\rho^{PSL}$, Inv$(\rho^{PSL})$ is a subgroup of $H^{1}(M,\mathbb{Z}_{2})$ defined as
\begin{equation}
\label{eq:inv}
    \text{Inv}(\rho^{PSL})\coloneq\{\eta\in H^{1}(M,\mathbb{Z}_{2})\,:\,\rho^{PSL}\otimes\eta =\rho^{PSL}\otimes 1\}\,.
\end{equation}
The Chern-Simons invariant $\text{CS}[\rho]$ and the adjoint Reidemeister torsion $\text{Tor}[\rho]$ are topological invariants of the flat connection. The CS invariant is defined as
\begin{equation}
    \text{CS}[\rho]\coloneq\frac{1}{8\pi^{2}}\int\Tr\left(\mathcal{A}_{\rho}d\mathcal{A}_{\rho}+\frac{2}{3}\mathcal{A}_{\rho}^{3}\right)\,.
\end{equation}
$\text{CS}[\rho]$ is defined modulo 1 for the $SL(2,\mathbb{C})$ connections, and modulo $\frac{1}{2}$ for the $PSL(2,\mathbb{C})$ connections. $\text{Tor}[\rho]$ corresponds to the 1-loop part of perturbative expansion of $SL(2,\mathbb{C})$ CS theory around the flat connection $\mathcal{A}_{\rho}$ \cite{Witten:1988hf}. Also, note that in Table~\ref{tab:bosdict}, for $\rho^{PSL}$ with nontrivial 2nd Stiefel-Whitney class $w_{2}(\rho^{PSL})$, conformal dimension of $\rho^{PSL}\otimes\eta$ is given only modulo $\frac{1}{2}$ due to the lack of our knowledge. For more details, see \cite{Cho:2020ljj}.

Our main conjecture in this section is given as follows.

\paragraph{\boldmath Conjecture}
\begin{align}\label{conj}
    \mathcal{T}_{SM(P, Q)}\simeq T_{\text{irred}}[S^2((P,P-R),(Q,S),(3,1))]
\end{align}
, where $PS-QR=2$ and $\gcd(P,R)=\gcd(Q,S)=1$.
Let us first explain the equivalence relations \cite{Gang:2024}:
\begin{equation}
    \begin{aligned}
        &\mathcal{T}_{1}\sim\mathcal{T}_{2}\text{ if two theories are IR equivalent up to some `topological operations',}\\
        &\mathcal{T}_{1}\simeq\mathcal{T}_{2}\text{ if two theories are IR equivalent up to some `minimal topological operations'.}
    \end{aligned}
\end{equation}
Examples of topological operations include tensoring with a unitary TQFT, gauging of finite (or generalized) symmetries, time-reversal, and so on. Among them, the minimal topological operations are the ones that preserve the absolute values of partition functions on arbitrary closed 3-manifolds, such as tensoring with an invertible TQFT, time-reversal, and so on. $\simeq$ is a stronger equivalence than $\sim$. Note that the condition $PS-QR=2$ fixes $(R,S)$ up to the shift $(R,S)\rightarrow(R,S)+\mathbb{Z}(P, Q)$. As in the previous paper\cite{Gang:2024}, we claim that $T_{\text{irred}}[S^2((P,P-R),(Q,S),(3,1))]$ is independent of the shift:
\begin{equation}
\label{eq:shift}
    \begin{aligned}
        &T_{\text{irred}}\left[S^2((P,P-R),(Q,S),(3,1))\right]\simeq T_{\text{irred}}\left[S^2((P,P-\tilde{R}),(Q,\tilde{S}),(3,1))\right],\\
        &\text{where }(\tilde{R},\tilde{S})=(R,S)+n(P, Q)\text{ for arbitrary }n\in\mathbb{Z}\,.
    \end{aligned}
\end{equation}
Also, the $\mathcal{T}_{SM(P, Q)}$, like $SM(P, Q)$, is invariant under the exchange of $P\leftrightarrow Q$:
\begin{equation}
    \begin{aligned}
        \mathcal{T}_{SM(Q,P)}&\simeq T_{\text{irred}}\left[S^2((Q,Q-\tilde{R}),(P,\tilde{S}),(3,1))\right]\text{ with }Q\tilde{S}-P\tilde{R}=2\\
        &\simeq T_{\text{irred}}\left[S^2((P,(\tilde{S}-P)+P),(Q,Q-\tilde{R}),(3,1))\right]\simeq \mathcal{T}_{SM(P, Q)}\,.
    \end{aligned}
\end{equation}
In the second line, we used the fact that $P(Q-\tilde{R})-Q(P-\tilde{S})=2$. Note that for bosonic theories, only $\widetilde{\mathcal{T}}_{SM(P, Q)}\sim\widetilde{T}_{\text{irred}}[M]$ may hold. For example, consider some $\mathcal{T}_{1},\,\mathcal{T}_{2}$ and their bosonization $\widetilde{\mathcal{T}}_{1},\,\widetilde{\mathcal{T}}_{2}$ such that
\begin{equation}
    \begin{aligned}
        &\mathcal{T}_{1}=\mathcal{T}_{2}\otimes(\text{Free Majorana-Weyl fermion})\,,\\
        &\widetilde{\mathcal{T}}_{1}=\frac{\widetilde{\mathcal{T}}_{2}\otimes(\text{Ising TQFT})}{\mathbb{Z}_{2}^{\text{diag}}}\,.
    \end{aligned}
\end{equation}
We can see that $\widetilde{\mathcal{T}}_{1}$ and $\widetilde{\mathcal{T}}_{2}$ are not related by minimal topological operations even if $\mathcal{T}_{1}\simeq\mathcal{T}_{2}$ holds. In this paper we will focus on the relation $\mathcal{T}_{SM(P, Q)}\simeq T_{\text{irred}}[M]$, and leave the computation about $\widetilde{\mathcal{T}}_{SM(P, Q)}\sim\widetilde{T}_{\text{irred}}[M]$ for the future work.
To check the conjecture \eqref{conj}, we identify the simple objects (irreducible PSL$(2,\mathds{C})$ flat connection) and modular data of the theory $T_{\text{irred}}[S^2((P,P-R),(Q,S),(3,1))]$.

\paragraph
{\boldmath Irreducible PSL$(2,\mathds{C})$ flat connections}
Let us check the proposal \eqref{conj} using the dictionaries Table~\ref{tab:ferdict} and \ref{tab:bosdict}. The fundamental group of the Seifert fibered manifold $S^{2}(\vec{p},\vec{q})$ can be written as
\begin{equation}
    \pi_{1}(S^{2}(\vec{p},\vec{q}))=\langle x_{1},x_{2},x_{3},h|x_{i}^{p_{i}}h^{q_{i}}=1,x_{1}x_{2}x_{3}=1,h\text{ is central}\rangle\,.
\end{equation}
Then, irreducible $PSL(2,\mathbb{C})$ flat connections satisfying $x^{p_1}=y^{p_2}=(xy)^{p_3}=1\,(h\,\text{trivial})$ can be written as \cite{Abdelghani:2001}
\begin{equation}
\begin{aligned}
    &\rho_{j}(x)=\begin{pmatrix}
\alpha_{j}&0\\0&\alpha_{j}^{-1}
\end{pmatrix}\,,\\
&\rho_{k}(y)=\begin{pmatrix}
\gamma&1\\ \gamma(\beta-\gamma)-1&\beta-\gamma
\end{pmatrix}\sim
\begin{pmatrix}
    e^{\frac{\pi ik}{p_{2}}}&0\\0&e^{-\frac{\pi ik}{p_{2}}}
\end{pmatrix}\,,\\
&\rho((xy)^{-1})\sim
\begin{pmatrix}
    e^{\frac{\pi il}{p_{3}}}&0\\0&e^{-\frac{\pi il}{p_{3}}}
\end{pmatrix}\,,\\
&\alpha_{j}=e^{\frac{\pi ij}{p_{1}}}\,,\,\beta=2\cos{\frac{\pi k}{p_{2}}}\,,\,\gamma=\frac{2\cos{\frac{\pi l}{p_{3}}}-\alpha_{j}^{-1}\beta}{2i\text{Im}\alpha_{j}}\,.
\end{aligned}
\end{equation}
Here $\sim$ means the equivalence up to the similarity transformation. In our case, for $S^{2}((P,P-R),(Q,S),(3,1)),$ we will take $p_{1}=3,\,p_{2}=P,\,p_{3}=Q\,.$ Distinct irreducible connections correspond to
\begin{equation}
\label{eq:jkls}
    \begin{aligned}
        &P,\,Q\,\text{odd}:\,j=1,\,k=1,\cdots,\lfloor\frac{P}{2}\rfloor,\,l=1,\cdots,Q-1\,,\\
        &P,\,Q\,\text{even}:\,j=1,\,k=1,\cdots,\lfloor\frac{P}{2}\rfloor-1,\,l=1,\cdots,Q-1\,\,\text{and}\,\,k=\frac{P}{2},\,l=1,\cdots,\frac{Q}{2}\,.
    \end{aligned}
\end{equation}
For each flat connection, let us define $(n_{1},n_{2},n_{3},\lambda)$ as follows:
\begin{equation}
    \begin{aligned}
        &\rho(h)=\text{diag}\{e^{2\pi i\lambda},e^{-2\pi i\lambda}\}\text{ with }\lambda_{\alpha}\in\{0,\frac{1}{2}\}\,,\\
        &\text{eigenvalues of }\rho(x_{i})=\left\{\exp\left(\pm 2\pi i\frac{n_{i}}{p_{i}}\right)\right\}\text{ with }n_{i}\in\frac{1}{2}\mathbb{Z}\,.
    \end{aligned}
\end{equation}
In all, irreducible flat connections are characterized by $(n_{1},n_{2},n_{3},\lambda)=(\frac{j}{2},\frac{k}{2},\frac{l}{2},0)$, where $n_{i},\lambda$ are modulo $\frac{p_{i}}{2},\frac{1}{2}$. In our case \eqref{eq:jkls}, $j=1$ and the connections depend on $k$ and $l$. Also, it can be checked that for $P, Q$ odd(even), connections with odd(even) $k+l$ has trivial $w_{2}[\rho^{PSL}]$ and can be lifted to the $SL(2,\mathbb{C})$ connections. Note that when we lift $\rho^{PSL}$ to the $SL(2,\mathbb{C})$ connection, we should specify $n_{i}$ (mod $p_{i}$), $\lambda$ (mod 1). Chern-Simons invariant and adjoint Reidemeister torsion of the $SL(2,\mathbb{C})$ connection is
\begin{equation}
    \begin{aligned}
        &CS[\rho]=\sum_{i=1}^{3}\left(\frac{r_{i}}{p_{i}}n_{i}^{2}-q_{i}s_{i}\lambda^{2}\right)\,,\\
        &\text{Tor}[\rho]=\prod_{i=1}^{3}\frac{p_{i}}{4\sin^{2}\left(2\pi(\frac{r_{i}}{p_{i}}n_{i}+s_{i}\lambda)\right)}\,,
    \end{aligned}
\end{equation}
where the integers $(r_{i},s_{i})$ are chosen such that $p_{i}s_{i}-q_{i}r_{i}=1$.

We propose the following one-to-one map between the primaries $\mathcal{O}^{\text{NS}}_{(a,b)}$ of $SM(P, Q)$ and the irreducible $PSL(2,\mathbb{C})$ connections $\rho_{(k,l)}$ on $S^{2}((P,P-R),(Q,S),(3,1))$ with trivial $w_{2}[\rho_{(k,l)}]$:
\begin{equation}
\label{eq:chprmap}
    \begin{aligned}
        &\mathcal{O}^{\text{NS}}_{(a,b)}\leftrightarrow\rho_{(k,l)}\,,\\
        &\text{where } (a,b)=\begin{cases}
            (k,Q-l)\,,&P, Q\text{ odd}\\
            (k,l)\,,&P, Q\text{ even}
        \end{cases}\,.
    \end{aligned}
\end{equation}
Under the map, one can check that $(\text{Tor}(\rho_{\alpha}))^{-1}$(resp. $CS[\rho_{\alpha=0}]-CS[\rho_{\alpha}]$) equals to $(S^{\text{NS}}_{0\alpha})^{2}$(resp. $h_{\alpha}\pmod{\frac{1}{2}}$) of $SM(P, Q)$. First, note that for every possible $P, Q$,
\begin{equation}
    \text{Tor}[\rho_{(k,l)}]=\frac{P}{4\sin^{2}\left(\frac{\pi r_{1}k}{P}\right)}\frac{Q}{4\sin^{2}\left(\frac{\pi r_{2}l}{Q}\right)}\,,
\end{equation}
where $Ps_{1}-(P-R)r_{1}=1,\,Qs_{2}-Sr_{2}=1\,.$ One can check that
\begin{equation}
    \begin{aligned}
        &\frac{Pb-Qa}{2}\equiv r_{1}k\pmod{P}\Leftrightarrow a\equiv k\pmod{P},\\
        &\frac{Pb-Qa}{2}\equiv r_{2}l\pmod{Q}\Leftrightarrow -b\equiv l\pmod{Q}\,,
    \end{aligned}
    \label{eq:mod}
\end{equation}
which implies
\begin{equation}
    (S^{\text{NS}}_{(1,1),(a,b)})^2 =\begin{cases}
    (\text{Tor}[\rho_{(k=a,l=Q-b)}])^{-1}\,,&P, Q\text{ odd}\\
    (\text{Tor}[\rho_{(k=a,l=b)}])^{-1}\,,&P, Q\text{ even}
    \end{cases}\,.
\end{equation}
Equality of $CS[\rho_{\alpha=0}]-CS[\rho_{\alpha}]$ and $h_{\alpha}\pmod{\frac{1}{2}}$ can be checked experimentally. Let us give some examples.
\paragraph{Example: $SM(2,8)$ from $S^{2}((2,1),(8,5),(3,1))$}
We choose $(R,S)=(1,5)$. There are two irreducible $PSL(2,\mathbb{C})$ connections $\rho_{(j,k,l,\lambda)}$ with trivial $w_{2}[\rho]$:
\begin{table}[H]
    \centering
    \begin{tabular}{|c|c|c|c|c|c|}
    \hline
        $(j,k,l)$&$(\vec{n},\lambda)$ & $\mathcal{O}^{\text{NS}}_{(a,b)}$ & $CS[\rho]$& Tor$[\rho]$ & $CS[\rho_{\alpha=0}]-CS[\rho]$ \\
        \hline
        $(1,1,1)$&$(\frac{1}{2},\frac{1}{2},\frac{1}{2},\frac{1}{2})$ & $\mathcal{O}^{\text{NS}}_{(1,1)}=\mathcal{O}^{\text{NS}}_{(1,7)}$ & $\frac{37}{96}$ & $\frac{4}{2+\sqrt{2}}$ & $0$ \\
        \hline
        $(1,1,3)$&$(\frac{1}{2},\frac{1}{2},\frac{3}{2},\frac{1}{2})$ & $\mathcal{O}^{\text{NS}}_{(1,3)}=\mathcal{O}^{\text{NS}}_{(1,5)}$ & $\frac{13}{96}$ & $\frac{4}{2-\sqrt{2}}$ &$\frac{1}{4}$ \\
        \hline
    \end{tabular}
    \caption{$SM(2,8)$ from $S^2 ((2,1),(8,5),(3,1))$. CS invariant is defined modulo $\frac{1}{2}$. We use the characters-to-primaries map in \eqref{eq:chprmap}. The result is compatible with that $S^{\text{NS}}_{(1,1),(1,1)}=\frac{\sqrt{2+\sqrt{2}}}{2}\,,\,S^{\text{NS}}_{(1,1),(1,3)}=\frac{\sqrt{2-\sqrt{2}}}{2}$ and $h_{(1,3)}=-\frac{1}{4}$.}
    \label{tab:SM28}
\end{table}
\paragraph{Example: $SM(3,5)$ from $S^{2}((3,4),(5,-1),(3,1))$}
We choose $(R,S)=(-1,-1)$. There are two irreducible $PSL(2,\mathbb{C})$ connections $\rho_{(j,k,l,\lambda)}$ with trivial $w_{2}[\rho]$:
\begin{table}[H]
    \centering
    \begin{tabular}{|c|c|c|c|c|c|}
    \hline
        $(j,k,l)$&$(\vec{n},\lambda)$ & $\mathcal{O}^{\text{NS}}_{(a,b)}$ & $CS[\rho]$& Tor$[\rho]$ & $CS[\rho_{\alpha=0}]-CS[\rho]$ \\
        \hline
        $(1,1,2)$&$(2,2,1,0)$ & $\mathcal{O}^{\text{NS}}_{(1,3)}=\mathcal{O}^{\text{NS}}_{(2,2)}$ & $\frac{8}{15}$ & $\frac{5-\sqrt{5}}{2}$ & $-\frac{2}{5}$ \\
        \hline
        $(1,1,4)$&$(2,2,2,0)$ & $\mathcal{O}^{\text{NS}}_{(1,1)}=\mathcal{O}^{\text{NS}}_{(2,4)}$ & $\frac{2}{15}$ & $\frac{5+\sqrt{5}}{2}$ &$0$ \\
        \hline
    \end{tabular}
    \caption{$SM(3,5)$ from $S^{2}((3,4),(5,-1),(3,1))$. CS invariant is defined modulo $\frac{1}{2}$. The result is compatible with that $S^{\text{NS}}_{(1,1),(1,1)}=\sqrt{\frac{5-\sqrt{5}}{10}}\,,\,S^{\text{NS}}_{(1,1),(1,3)}=\sqrt{\frac{5+\sqrt{5}}{10}}$ and $h_{(1,3)}=\frac{1}{10}$.}
    \label{tab:SM35}
\end{table}
\paragraph{Example: $SM(3,7)$ from $S^{2}((3,2),(7,3),(3,1))$}
We choose $(R,S)=(1,3)$. There are three irreducible $PSL(2,\mathbb{C})$ connections $\rho_{(j,k,l,\lambda)}$ with trivial $w_{2}[\rho]$:
\begin{table}[H]
    \centering
    \begin{tabular}{|c|c|c|c|c|c|}
    \hline
        $(j,k,l)$&$(\vec{n},\lambda)$ & $\mathcal{O}^{\text{NS}}_{(a,b)}$ & $CS[\rho]$& Tor$[\rho]$ & $CS[\rho_{\alpha=0}]-CS[\rho]$ \\
        \hline
        $(1,1,2)$&$(2,2,1,0)$ & $\mathcal{O}^{\text{NS}}_{(1,5)}=\mathcal{O}^{\text{NS}}_{(2,2)}$ & $\frac{2}{7}$ & $\frac{7}{4}\sec ^{2}\frac{\pi}{14}$ & $\frac{2}{7}$ \\
        \hline
        $(1,1,4)$&$(2,2,2,0)$ & $\mathcal{O}^{\text{NS}}_{(1,3)}=\mathcal{O}^{\text{NS}}_{(2,4)}$ & $\frac{1}{7}$ & $\frac{7}{4}\csc ^{2}\frac{\pi}{7}$ &$\frac{3}{7}$ \\
        \hline
        $(1,1,6)$&$(2,2,3,0)$ & $\mathcal{O}^{\text{NS}}_{(1,1)}=\mathcal{O}^{\text{NS}}_{(2,6)}$ & $\frac{4}{7}$ & $\frac{7}{4}\sec ^{2}\frac{3\pi}{14}$ &$0$ \\
        \hline
    \end{tabular}
    \caption{$SM(3,7)$ from $S^{2}((3,2),(7,3),(3,1))$. CS invariant is defined modulo $\frac{1}{2}$. The result is compatible with that $S^{\text{NS}}_{(1,1),(1,1)}=\frac{2}{\sqrt{7}}\cos\frac{3\pi}{14}\,,\,S^{\text{NS}}_{(1,1),(2,2)}=-\frac{2}{\sqrt{7}}\cos\frac{\pi}{14}\,,\,S^{\text{NS}}_{(1,1),(2,4)}=\frac{2}{\sqrt{7}}\sin\frac{\pi}{7}$ and $h_{(2,2)}=\frac{2}{7}\,,\,h_{(2,4)}=-\frac{1}{14}$.}
    \label{tab:SM28}
\end{table}
\paragraph{Example: $SM(4,10)$ from $S^{2}((4,3),(10,3),(3,1))$}
We choose $(R,S)=(1,3)$. There are 7 irreducible $PSL(2,\mathbb{C})$ connections $\rho_{(j,k,l,\lambda)}$ with trivial $w_{2}[\rho]$:
\begin{table}[H]
    \centering
    \begin{tabular}{|c|c|c|c|c|c|}
    \hline
        $(j,k,l)$&$(\vec{n},\lambda)$ & $\mathcal{O}^{\text{NS}}_{(a,b)}$ & $CS[\rho]$& Tor$[\rho]$ & $CS[\rho_{\alpha=0}]-CS[\rho]$ \\
        \hline
        $(1,1,1)$&$(\frac{1}{2},\frac{1}{2},\frac{1}{2},\frac{1}{2})$ & $\mathcal{O}^{\text{NS}}_{(1,1)}=\mathcal{O}^{\text{NS}}_{(3,9)}$ & $\frac{133}{240}$ & $\frac{80}{(1+\sqrt{5})^{2}}$ & $0$ \\
        \hline
        $(1,1,3)$&$(\frac{1}{2},\frac{1}{2},\frac{3}{2},\frac{1}{2})$ & $\mathcal{O}^{\text{NS}}_{(1,3)}=\mathcal{O}^{\text{NS}}_{(3,7)}$ & $\frac{37}{240}$ & $\frac{80}{(-1+\sqrt{5})^{2}}$ &$\frac{2}{5}$ \\
        \hline
        $(1,1,5)$&$(\frac{1}{2},\frac{1}{2},\frac{5}{2},\frac{1}{2})$ & $\mathcal{O}^{\text{NS}}_{(1,5)}=\mathcal{O}^{\text{NS}}_{(3,5)}$ & $\frac{17}{48}$ & $5$ &$\frac{1}{5}$ \\
        \hline
        $(1,1,7)$&$(\frac{1}{2},\frac{1}{2},\frac{7}{2},\frac{1}{2})$ & $\mathcal{O}^{\text{NS}}_{(1,7)}=\mathcal{O}^{\text{NS}}_{(3,3)}$ & $\frac{37}{240}$ & $\frac{80}{(-1+\sqrt{5})^{2}}$ &$\frac{2}{5}$ \\
        \hline
        $(1,1,9)$&$(\frac{1}{2},\frac{1}{2},\frac{9}{2},\frac{1}{2})$ & $\mathcal{O}^{\text{NS}}_{(1,9)}=\mathcal{O}^{\text{NS}}_{(3,1)}$ & $\frac{133}{240}$ & $\frac{80}{(1+\sqrt{5})^{2}}$ &$0$ \\
        \hline
        $(1,2,2)$&$(2,3,1,0)$ & $\mathcal{O}^{\text{NS}}_{(2,2)}=\mathcal{O}^{\text{NS}}_{(2,8)}$ & $\frac{13}{60}$ & $5-\sqrt{5}$ &$\frac{27}{80}$ \\
        \hline
        $(1,2,4)$&$(2,3,2,0)$ & $\mathcal{O}^{\text{NS}}_{(2,4)}=\mathcal{O}^{\text{NS}}_{(2,6)}$ & $\frac{7}{60}$ & $5+\sqrt{5}$ &$\frac{7}{16}$ \\
        \hline
    \end{tabular}
    \caption{$SM(4,10)$ from $S^{2}((4,3),(10,3),(3,1))$. CS invariant is defined modulo $\frac{1}{2}$. The result is compatible with the NS sector modular data of the $SM(4,10)$, which can be computed from \eqref{eq:smsmat} and \eqref{eq:smtmat}.}
    \label{tab:SM410}
\end{table}
Also, for the examples above, it can be checked that for $\eta\neq 1$, $CS[\rho_{\alpha}\otimes 1]-CS[\rho_{\alpha}\otimes\eta]=\frac{1}{2}\pmod{1}$. Accordingly, we can indeed see that $\mathcal{T}_{SM(P, Q)}$ is a fermionic theory.

\subsection{Field theory description of $\mathcal{T}_{SM(P, Q)}$}
In this section, we propose a field theory description of $\mathcal{T}_{SM(P, Q)}$ using $T[SU(2)]$ theory. In the previous work \cite{Gang:2024}, the field theory description of $T_{\text{irred}}[S^{2}(\Vec{p},\Vec{q})]$ based on the Dehn surgery prescription \cite{Gang:2018wek,Gukov:2017kmk,Assel:2022row,Pei:2015jsa,Alday:2017} was introduced. We will briefly review those results. For more details, refer to \cite{Gang:2024}. In \cite{Gang:2024}, it was proposed that
\begin{equation}
\label{eq:tirred}
    \begin{aligned}
        &T_{\text{irred}}\left[M=S^{2}((p_{1},q_{1}),(p_{2},q_{2}),(p_{3},q_{3}))\right]\otimes(\text{a unitary TQFT})\\
        &\simeq [(\mathcal{D}(p_{1},q_{1})\otimes\mathcal{D}(p_{2},q_{2})\otimes\mathcal{D}(p_{3},q_{3}))]/H^{1}(M,\mathbb{Z}_{2})\,.
    \end{aligned}
\end{equation}
Here $H^{1}(M,\mathbb{Z}_{2})$ denotes the 1-form symmetry in $\prod_{i}\mathcal{D}(p_{i},q_{i})$ which geometrically originates from the $\mathbb{Z}_{2}$ cohomology of the internal 3-manifold in 3D-3D correspondence \cite{Eckhard:2019jgg,Cho:2020ljj}. $\mathcal{D}(P, Q)$ is defined as follows:
\begin{equation}
\label{eq:calddef}
    \begin{aligned}
        &D(\Vec{k})\simeq\mathcal{D}(P, Q)\otimes\text{TFT}[\Vec{k}],\text{ where}\\
        &D(\Vec{k})\coloneq\begin{cases}
            \frac{T[SU(2)]^{\otimes(\sharp-1)}}{SU(2)^{(1)}_{k^{(1)}}\times SU(2)^{(2)}_{k^{(2)}}\dots\times SU(2)^{(\sharp)}_{k^{(\sharp)}}}, & \sharp\geq 2\\
            \mathcal{N}=2\text{ pure }SU(2)_{k^{(1)}}\text{ CS theory}, & \sharp=1
        \end{cases}\,.
    \end{aligned}
\end{equation}
Here $/G_{k}$ denotes $\mathcal{N}=3$ gauging of $G$ symmetry with Chern-Simons level $k$. The CS levels $\vec{k}=(k^{(1)},\dots,k^{(\sharp)})$ are related to the $(P, Q)$ as
\begin{equation}
\label{eq:dehnslope}
    \frac{Q}{P}=\frac{1}{k^{(1)}-\frac{1}{k^{(2)}-\frac{1}{k^{(3)}-\dots\frac{1}{k^{(\sharp)}}}}}\,.
\end{equation}
The gauged $SU(2)$ symmetries are
\begin{equation}
\begin{aligned}
    &SU(2)^{(1)}:SU(2)_{L}^{(1)}\coloneq SU(2)_{L}\text{ of the 1st }T[SU(2)]\,,\\
    &SU(2)^{(2\leq I\leq\sharp -1)}:\text{diagonal subgroup of }(SU(2)_{R}^{(I-1)}\times SU(2)_{L}^{(I)})\,,\\
    &SU(2)^{(\sharp)}:SU(2)_{R}^{(\sharp -1)}\,.
\end{aligned}
\end{equation}
The decoupled topological theory is given by
\begin{equation}
\begin{aligned}
    &\text{TFT}[\vec{k}=(k^{(1)},k^{(2)},\dots,k^{(\sharp)})]\\
    &=U(1)_{\mathcal{K}}^{\sharp}\text{ theory with mixed CS level }\mathcal{K}_{IJ}=2\times\begin{cases}
        +1\text{ or }-1\,, & |I-J|=1\\
        0\,, & I=J\text{ and }k^{(I)}\in 2\mathbb{Z}\\
        +1\text{ or }-1\,, & I=J\text{ and }k^{(I)}\in 2\mathbb{Z}+1\\
        0\,, & |I-J|>0
    \end{cases}\,.
\end{aligned}
\end{equation}
In \cite{Gang:2024}, properties of $\mathcal{D}(P, Q)$ were also analyzed. It was shown that
\begin{equation}
\label{eq:dproperties}
    \begin{aligned}
        &i)\,\mathcal{D}(P, Q)\text{ does not depend on the choice of }\vec{k}\text{ for given }(P, Q)\,,\\
        &ii)\,\mathcal{D}(P, Q)\simeq\mathcal{D}(P, Q+P\mathbb{Z})\,.
    \end{aligned}
\end{equation}
Also, it was argued that the IR phase of the $\mathcal{D}(P, Q)$ theory is given as follows:
\begin{equation}
    \mathcal{D}(P, Q)\xrightarrow{\text{IR}}\begin{cases}
        \text{mass gap, unitary TQFT,} & Q=\pm 1\pmod{P}\\
        \mathcal{N}=4\text{ rank-0 SCFT,} & Q\neq\pm 1\pmod{P}
    \end{cases}\,.
\end{equation}
Note that,
\begin{equation}
\label{eq:unitaryd}
\begin{aligned}
    &\mathcal{D}(P,\pm1+P\mathbb{Z})\simeq\mathcal{D}(P,\pm1)\sim D(\vec{k}=(\pm P))=(\mathcal{N}=3\,\,SU(2)_{\pm P})\\
    &\simeq(\mathcal{N}=2\,\,SU(2)_{\pm P})\simeq SU(2)_{\pm P-2\times\text{sign}(\pm P)}\,.
\end{aligned}
\end{equation}
The pure $\mathcal{N}=3$ CS theory with non-zero CS level is IR equivalent to pure $\mathcal{N}=2$ CS theory with the same level, since the adjoint chiral multiplet in the $\mathcal{N}=3$ multiplet has a superpotential mass term and can be integrated out. The pure $\mathcal{N}=2$ CS theory $SU(2)_{k}$ contains an auxiliary massive gaugino, and integrating it out induces a CS level shift by $-2\times\text{sign}(k)$ \cite{Gang:2024}. From \eqref{eq:unitaryd}, we can see that\footnote{The first term, $SU(2)_{2-2}=SU(2)_{0}$ is a trivial theory. The second term, for the plus sign, $SU(2)_{3-2}=SU(2)_{1}\simeq U(1)_{2}$, and TFT$[\vec{k}=(3)]$ is also $U(1)_{2}$. The remaining theory is trivial \cite{Gang:2024}.}
\begin{equation}
\label{eq:trivial}
    \mathcal{D}(2,1+2\mathbb{Z})\simeq\mathcal{D}(3,\pm1+3\mathbb{Z})\simeq(\text{a trivial theory})\,.
\end{equation}

Using the above proposal \eqref{eq:tirred}, for the $\mathcal{T}_{SM(P, Q)}$ theory, we propose that
\begin{equation}
    \begin{aligned}
        &\mathcal{T}_{SM(P, Q)}\simeq T_{\text{irred}}[S^{2}((P,P-R),(Q,S),(3,1))]\\
        &\simeq
        \begin{cases}
        \mathcal{D}(P,P-R)\otimes\mathcal{D}(Q,S), & \text{if }P, Q\in 2\mathbb{Z}+1\\
        \left(\frac{\mathcal{D}(P,P-R)\otimes\mathcal{D}(Q,S)\otimes SU(2)_{2}}{\mathbb{Z}_{2}^{\text{diag}}}\right)/\mathbb{Z}_{2}\,, & \text{if }P, Q\in 2\mathbb{Z}
        \end{cases}\,.
    \end{aligned}
    \label{eq:fielddescription}
\end{equation}
This agrees with the three-sphere partition function computation:
\begin{equation}
    \begin{aligned}
        &|\mathcal{Z}_{b=1}^{\text{con}}\text{ of }\mathcal{T}_{SM(P, Q)}|\\
        &=
        \begin{cases}
        |(\mathcal{Z}^{\text{con}}_{b=1}\text{ of }\mathcal{D}(P,P-R))\times(\mathcal{Z}^{\text{con}}_{b=1}\text{ of }\mathcal{D}(Q,S))|, & \text{if }P, Q\in 2\mathbb{Z}+1\\
        |(\mathcal{Z}^{\text{con}}_{b=1}\text{ of }\mathcal{D}(P,P-R))\times(\mathcal{Z}^{\text{con}}_{b=1}\text{ of }\mathcal{D}(Q,S))\times\frac{1}{2}\times 2\times 2|, & \text{if }P, Q\in 2\mathbb{Z}
        \end{cases}\\
        &=\frac{8}{\sqrt{PQ}}\sin{\frac{\pi}{P}}\sin{\frac{\pi}{Q}}=\min |S_{0\alpha}^{NS}|\,.
    \end{aligned}
\end{equation}
We used the fact that the $S^{3}$ partition function of $SU(2)_{2}$ is $\frac{1}{2}$. Note that only for even $P$, $\mathcal{D}(P, Q)$ has a non-anomalous $\mathbb{Z}_2$ one-form symmetry.\cite{Gang:2024}

When $Q=P+2$, the $(R,S)$ can be chosen as $(-1,-1)$ and the $\mathcal{T}_{SM(P, Q)}$ theory becomes
\begin{equation}
\label{eq:unitaryfield}
    \begin{aligned}
    &
    \begin{cases}
        \mathcal{D}(P,P+1)\otimes\mathcal{D}(P+2,-1),& \text{if }P, Q\in 2\mathbb{Z}+1\\
        \left(\frac{\mathcal{D}(P,P+1)\otimes\mathcal{D}(P+2,-1)\otimes SU(2)_{2}}{\mathbb{Z}_{2}^{\text{diag}}}\right)/\mathbb{Z}_{2},& \text{if }P, Q\in 2\mathbb{Z}
    \end{cases}\\
    &\simeq\left(\frac{SU(2)_{(P-2)}\otimes SU(2)_{(-P)}\otimes SU(2)_{2}}{\mathbb{Z}_{2}^{\text{diag}}}\right)/\mathbb{Z}_{2}\,.
    \end{aligned}
\end{equation}
This is the coset description of the 3d TQFT corresponding to the unitary supersymmetric minimal model $\mathcal{T}_{SM(P, P+2)}$ \cite{Moore:1989}. Note that in \eqref{eq:unitaryfield}, the equivalence for $P, Q\in 2\mathbb{Z}+1$ case follows from
\begin{equation}
\label{eq:unitaryfield2}
    \begin{aligned}
        &\left(\frac{SU(2)_{(P-2)}\otimes SU(2)_{(-P)}\otimes SU(2)_{2}}{\mathbb{Z}_{2}^{\text{diag}}}\right)/\mathbb{Z}_{2}\\
        &\simeq\mathcal{D}(P,P+1)\otimes\mathcal{D}(P+2,-1)\otimes\left(\frac{U(1)_{\pm2}\otimes U(1)_{\pm2}\otimes SU(2)_{2}}{\mathbb{Z}_{2}^{\text{diag}}}\right)/\mathbb{Z}_{2}\\
        &\simeq\mathcal{D}(P,P+1)\otimes\mathcal{D}(P+2,-1)\,.
    \end{aligned}
\end{equation}
In \eqref{eq:unitaryfield} and \eqref{eq:unitaryfield2} we used the fact that
\begin{equation}
    \begin{aligned}
        \text{For }P\in 2\mathbb{Z}_{\geq 1}&+1,\,\mathcal{D}(P+2,-1)\otimes U(1)_{\pm 2}\simeq SU(2)_{-P}\,,\\
        &\text{ and }\mathcal{D}(P,P+1)\otimes U(1)_{\pm 2}\simeq SU(2)_{(P-2)}\,,\\
        \text{For }P\in 2\mathbb{Z}_{\geq 1}&,\,\mathcal{D}(P+2,-1)\simeq SU(2)_{-P}\text{ and }\mathcal{D}(P,P+1)\simeq SU(2)_{(P-2)}\,.
    \end{aligned}
\end{equation}
\paragraph{Example : $(P, Q)=(3,7)$}Choosing $(R,S)=(1,3)$ and using equations \eqref{eq:unitaryfield} and \eqref{eq:trivial}, we find:
\begin{equation}
\begin{aligned}
    \mathcal{T}_{SM(3,7)}&\simeq \mathcal{D}(3,2)\otimes\mathcal{D}(7,3)\simeq\mathcal{D}(7,3)\,.
    \end{aligned}
\end{equation}
From that $\frac{3}{7}=\frac{1}{2-\frac{1}{-3}}$, we obtain:
\begin{equation}
    \mathcal{D}(7,3)\otimes\text{TFT}[\vec{k}]\simeq D(\vec{k})\text{ with }\vec{k}=(2,-3)\,.
\end{equation}
Let us define $\textbf{HF}_{\text{fer}}\coloneq \{(\mathcal{H}^{-1/2},(\mathcal{F}/\mathcal{F}_{\alpha =0})^{2})\}$ and $\textbf{HF}_{\text{bos}}\coloneq \{(\mathcal{H}^{-1/2},\mathcal{F}/\mathcal{F}_{\alpha =0})\}$ for fermionic and bosonic theories, respectively. Utilizing the explicit computation from \eqref{eq:hfcomputation}, we confirm that the set $\textbf{HF}_{\text{bos}}$ of the $D((2,-3))$ theory in the A-twisting limit can be factorized as follows:
\begin{equation}
    \begin{aligned}
        &\textbf{HF}_{\text{bos}}^{A}\text{ of }D((2,-3))\\
        &=\left\{\left(\frac{2}{\sqrt{7}}\cos{\frac{3\pi}{14}},1\right),\left(\frac{2}{\sqrt{7}}\cos{\frac{\pi}{14}},e^{\frac{4\pi i}{7}}\right),\left(\frac{2}{\sqrt{7}}\sin{\frac{\pi}{7}},e^{\frac{6\pi i}{7}}\right)\right\}\times\left\{\left(\frac{1}{2},1\right)^{\otimes 2},\left(\frac{1}{2},i\right),\left(\frac{1}{2},\frac{1}{i}\right)\right\}\,.
    \end{aligned}
\end{equation}
The second factor can be interpreted as the contribution from the decoupled $\text{TFT}[\vec{k}]$. The first factor gives the set $\textbf{HF}_{\text{fer}}^{A}$ for the $\mathcal{T}_{SM(3,7)}$ theory:
\begin{equation}
    \textbf{HF}_{\text{fer}}^{A}\text{ of }\mathcal{T}_{SM(3,7)}=
    \left\{\left(\frac{2}{\sqrt{7}}\cos{\frac{3\pi}{14}},1\right),\left(\frac{2}{\sqrt{7}}\cos{\frac{\pi}{14}},e^{\frac{8\pi i}{7}}\right),\left(\frac{2}{\sqrt{7}}\sin{\frac{\pi}{7}},e^{\frac{12\pi i}{7}}\right)\right\}\,.
\end{equation}
This set nicely matches with the set of $(|S_{0\alpha}^{\text{NS}}|,e^{4\pi ih_{\alpha}})$ for $SM(3,7)$, as expected from the dictionaries in Table~\ref{tab:ferdict}.
\paragraph{Example : $(P, Q)=(4,10)$}Choosing $(R,S)=(1,3)$, we have:
\begin{equation}
    \mathcal{T}_{SM(4,10)}\simeq\left(\frac{\mathcal{D}(4,3)\otimes\mathcal{D}(10,3)\otimes SU(2)_{2}}{\mathbb{Z}_{2}^{\text{diag}}}\right)/\mathbb{Z}_{2}\,.
\end{equation}
Using $\frac{3}{10}=\frac{1}{3-\frac{1}{-3}}$, we obtain:
\begin{equation}
    \mathcal{D}(10,3)\otimes\text{TFT}[\vec{k}]\simeq D(\vec{k})\text{ with }\vec{k}=(3,-3)\,.
\end{equation}
Using the explicit computation in \eqref{eq:hfcomputation}, we confirm that the set $\textbf{HF}_{\text{bos}}\coloneq \{(\mathcal{H}^{-1/2},\mathcal{F}/\mathcal{F}_{\alpha =0})\}$ of $D((3,-3))$ theory in the $A$-twisting limit can be factorized as follows:
\begin{equation}
    \begin{aligned}
        &\textbf{HF}_{\text{bos}}^{A}\text{ of }D((3,-3))\\
        &=\biggl\{\left(\frac{1}{\sqrt{5}},e^{\frac{2\pi i}{5}}\right),\left(\frac{1}{\sqrt{5}}\cos{\frac{\pi}{5}},1\right)^{\otimes 2},\left(\frac{1}{\sqrt{5}}\sin{\frac{\pi}{10}},e^{\frac{4\pi i}{5}}\right)^{\otimes 2},\left(\frac{1}{2\sqrt{2}}\sqrt{1-\frac{1}{\sqrt{5}}},e^{-\frac{\pi i}{4}}\right),\\
        &\left(\frac{1}{2\sqrt{2}}\sqrt{1-\frac{1}{\sqrt{5}}},e^{\frac{3\pi i}{4}}\right),\left(\frac{1}{2\sqrt{2}}\sqrt{1+\frac{1}{\sqrt{5}}},e^{-\frac{9\pi i}{20}}\right),\left(\frac{1}{2\sqrt{2}}\sqrt{1+\frac{1}{\sqrt{5}}},e^{\frac{11\pi i}{20}}\right)\biggr\}\times\left\{\left(\frac{1}{\sqrt{2}},1\right),\left(\frac{1}{\sqrt{2}},i\right)\right\}\,.
    \end{aligned}
\end{equation}
The second factor can be regarded as the contribution from the decoupled $\text{TFT}[\vec{k}]$, and the first factor is from $\mathcal{D}(10,3)$. From the $\textbf{HF}_{\text{bos}}^{A}$ of $\mathcal{D}(10,3)$, one can see that the $\mathcal{D}(10,3)$ theory has a non-anomalous $\mathbb{Z}_{2}$ 1-form symmetry generated by an anyon with topological spin 0. Hence, we have the set for $\mathcal{T}_{SM(4,10)}$ theory:
\begin{equation}
    \begin{aligned}
        &\textbf{HF}_{\text{fer}}^{A}\text{ of }\mathcal{T}_{SM(4,10)}\simeq\left(\frac{\mathcal{D}(4,3)\otimes\mathcal{D}(10,3)\otimes SU(2)_{2}}{\mathbb{Z}_{2}^{\text{diag}}}\right)/\mathbb{Z}_{2}\\
        &=\biggl\{\left(\frac{1}{\sqrt{5}},e^{\frac{4\pi i}{5}}\right),\left(\frac{1}{\sqrt{5}}\cos{\frac{\pi}{5}},1\right)^{\otimes 2},\left(\frac{1}{\sqrt{5}}\sin{\frac{\pi}{10}},e^{\frac{8\pi i}{5}}\right)^{\otimes 2},\left(\frac{1}{2}\sqrt{1-\frac{1}{\sqrt{5}}},e^{\frac{7\pi i}{4}}\right),\left(\frac{1}{2}\sqrt{1+\frac{1}{\sqrt{5}}},e^{-\frac{13\pi i}{20}}\right)\biggr\}\,.
    \end{aligned}
\end{equation}
This set nicely matches the set of $(|S_{0\alpha}^{\text{NS}}|,e^{4\pi ih_{\alpha}})$ of $SM(4,10)$, as expected from the dictionaries in Table~\ref{tab:ferdict}.

\subsection{Comparison with the $\mathcal{T}_{SM(P, Q)}$ by Baek-Gang}
Recently in \cite{Baek:2025}, the UV abelian bulk field theory description for $SM(2,4r),\,SM(3,6r-5),\text{ and }SM(3,6r-7)\,(r\geq 2)$ were proposed. We will denote those theories as $\mathcal{T}^{BG}_{SM(2,4r)}$, $\mathcal{T}^{BG}_{SM(3,6r-5)}$, and $\mathcal{T}^{BG}_{SM(3,6r-7)}$, respectively. In this section, we will claim that they are equivalent to the descriptions of $\mathcal{T}_{SM(P, Q)}$ in \eqref{conj}, and provide evidence by matching the superconformal index and 3-sphere partition function of both theories.

\subsubsection{$SM(2,4r)$}
\label{sec:bg24r}
For the case when $(P, Q)=(2,4r)$, the 3D $\mathcal{T}_{SM(P, Q)}$ is (we choose $R=1,\,S=2r+1$)
\begin{equation}
    \begin{aligned}
        &\mathcal{T}_{SM(2,4r)}\simeq T_{\text{irred}}[S^{2}((2,1),(4r,2r+1),(3,1))]\\
        &\simeq\left(\frac{\mathcal{D}(4r,2r+1)\otimes SU(2)_{2}}{\mathbb{Z}_{2}^{\text{diag}}}\right)/\mathbb{Z}_{2}\simeq\frac{\mathcal{D}(4r,2r+1)}{\mathbb{Z}_{2}}\otimes\frac{SU(2)_{2}}{\mathbb{Z}_{2}}\simeq\frac{\mathcal{D}(4r,2r+1)}{\mathbb{Z}_{2}}\,.
    \end{aligned}
\end{equation}
Here, we used the fact that both $\mathcal{D}(2,1)$ and $SU(2)_{2}/\mathbb{Z}_{2}$ are trivial theories. Using $\frac{2 r+1}{4r}=\frac{1}{2-\frac{1}{r-\frac{1}{-2}}}$, the $\mathcal{D}(4r,2r+1)$ is given as
\begin{equation}
    D(\vec{k})\simeq\mathcal{D}(4r,2r+1)\otimes\text{TFT}[\vec{k}]\text{ with }\vec{k}=(2,r,-2)\,.
\end{equation}
Meanwhile, in \cite{Baek:2025} it was proposed that
\begin{equation}
    \begin{aligned}
        &\mathcal{T}_{SM(2,4r)}\otimes(\text{Free fermion})=\mathcal{T}^{BG}_{SM(2,4r)}\,,\text{ where}\\
        &\mathcal{T}^{BG}_{SM(2,4r)}\coloneq SU(2)_{k=r}^{\frac{1}{2}\oplus\frac{1}{2}}\\
        &\coloneq SU(2)\text{ gauge theory coupled to a half hypermultiplet and a half twisted}\\
        &\text{hypermultiplet in the fundamental representations with Chern-Simons level }k=r\,.
    \end{aligned}
\end{equation}
We now claim that the two descriptions of $\mathcal{T}_{SM(P, Q)}$ are equivalent, namely
\begin{equation}
    \mathcal{T}_{SM(2,4r)}^{BG}\simeq\frac{\mathcal{D}(4r,2r+1)}{\mathbb{Z}_{2}}\,.
\end{equation}
We will check the equivalence by computing various partition functions. First, consider the superconformal index of $\mathcal{D}(4r,2r+1)/\mathbb{Z}_{2}$. The 't Hooft anomaly of the one-form symmetry in $D(\vec{k}=(2,r,-2))$ is as follows:\footnote{For our 3-manifold $M$, we can think of the 4-manifolds $X$ and $Y$ such that $\partial X=\partial Y=M$. $\mathcal{M}_{4}=X\cup\bar{Y}$ is a closed 4-manifold obtained by gluing $X$ and orientation reversed $Y$ along the common boundary $M$.}
\begin{equation}
    S_{\text{anomaly}}=\pi\int_{\mathcal{M}_{4}}\left(\sum_{I=1}^{3}k^{(I)}\frac{\mathcal{P}(w_{2}^{(I)})}{2}+\sum_{J=1}^{2}w_{2}^{(J)}\cup w_{2}^{(J+1)}\right)\pmod{2\pi}\,.
\end{equation}
Here $w_{2}^{(I)}\in H^{2}(\mathcal{M}_{4},\mathbb{Z}_{2})$ is the second Stiefel-Whitney class of $SO(3)^{(I)}=SU(2)^{(I)}/\mathbb{Z}_{2}$ bundle. $\mathcal{P}(w_{2})=w_{2}^{2}\pmod{2}$ is the Pontryagin square operation. We find that $\mathbb{Z}_{2}^{\text{diag}}\subset\mathbb{Z}_{2}^{(1)}\times\mathbb{Z}_{2}^{(3)}$ is anomaly free, and we will gauge $D(\vec{k}=(2,r,-2))$ using this symmetry. From \eqref{eq:dkindex} and \eqref{eq:sumrange}, we can confirm that
\paragraph{\boldmath \underline{$r=2$}}
\begin{equation}
    \begin{aligned}
        &\mathcal{I}^{\text{sci}}_{D(\vec{k}=(2,2,-2))/\mathbb{Z}_{2}}(\eta,\nu=0)=\mathcal{I}_{\mathcal{T}^{BG}_{SM(2,8)}}^{\text{sci}}(q;\eta,\nu=0)\\
        &=1-\sqrt{q}-\frac{\left(\eta ^2+\eta
   +1\right) q}{\eta }-\left(\eta
   +\frac{1}{\eta }+2\right)
   q^{3/2}-\frac{(\eta +1)^2
   q^2}{\eta }-q^{5/2} \\
   &+\left(\eta
   ^2+\frac{1}{\eta ^2}+\eta
   +\frac{1}{\eta }+1\right)
   q^3+\frac{(\eta +1)^2 \left(\eta
   ^2+1\right) q^{7/2}}{\eta
   ^2}+\left(\eta ^2+\frac{1}{\eta
   ^2}+3 \eta +\frac{3}{\eta
   }+3\right) q^4+O\left(q^{\frac{9}{2}}\right),
   \end{aligned}
\end{equation}
\paragraph{\boldmath \underline{$r=3$}} 
\begin{equation}
    \begin{aligned}
    &\mathcal{I}^{\text{sci}}_{D(\vec{k}=(2,3,-2))/\mathbb{Z}_{2}}(\eta,\nu=0)=\mathcal{I}_{\mathcal{T}^{BG}_{SM(2,12)}}^{\text{sci}}(q;\eta,\nu=0)\\
    &=1-\sqrt{q}-\frac{\left(\eta ^2+\eta
   +1\right) q}{\eta }-\left(\eta
   +\frac{1}{\eta }+2\right)
   q^{3/2}-q^2+\frac{(\eta +1)^2
   \left(\eta ^2+1\right)
   q^{5/2}}{\eta ^2} \\
   &+\frac{2
   \left(\eta ^2+\eta +1\right)^2
   q^3}{\eta ^2}+\frac{2 \left((\eta
   +1)^2 \left(\eta ^2+\eta
   +1\right)\right) q^{7/2}}{\eta
   ^2}+\left(\eta ^2+\frac{1}{\eta
   ^2}+7 \eta +\frac{7}{\eta
   }+9\right) q^4 \\
   &-\left(\eta
   ^3+\frac{1}{\eta ^3}-\eta
   ^2-\frac{1}{\eta ^2}-7 \eta
   -\frac{7}{\eta }-11\right)
   q^{9/2}+O\left(q^5\right),
   \end{aligned}
\end{equation}
\paragraph{\boldmath \underline{$r=4$}} 
\begin{equation}
    \begin{aligned}
        &\mathcal{I}^{\text{sci}}_{D(\vec{k}=(2,4,-2))/\mathbb{Z}_{2}}(\eta,\nu=0)=\mathcal{I}_{\mathcal{T}^{BG}_{SM(2,16)}}^{\text{sci}}(q;\eta,\nu=0)\\
        &=1-\sqrt{q}-\frac{\left(\eta ^2+\eta
   +1\right) q}{\eta }-\left(\eta
   +\frac{1}{\eta }+2\right)
   q^{3/2}-q^2+\frac{(\eta +1)^2
   \left(\eta ^2+1\right)
   q^{5/2}}{\eta ^2} \\
   &+\frac{2
   \left(\eta ^2+\eta +1\right)^2
   q^3}{\eta ^2}+\frac{\left(\eta
   ^4+5 \eta ^3+7 \eta ^2+5 \eta
   +1\right) q^{7/2}}{\eta
   ^2}-\frac{\left(\eta ^6+\eta ^5-4
   \eta ^4-6 \eta ^3-4 \eta ^2+\eta
   +1\right) q^4}{\eta ^3} \\
   &-\frac{2
   \left(\eta ^6+\eta ^5-\eta ^4-3
   \eta ^3-\eta ^2+\eta +1\right)
   q^{9/2}}{\eta
   ^3}+O\left(q^5\right),
    \end{aligned}
\end{equation}
\paragraph{\boldmath \underline{$r=5$}} 
\begin{equation}
    \begin{aligned}
        &\mathcal{I}^{\text{sci}}_{D(\vec{k}=(2,5,-2))/\mathbb{Z}_{2}}(\eta,\nu=0)=\mathcal{I}_{\mathcal{T}^{BG}_{SM(2,20)}}^{\text{sci}}(q;\eta,\nu=0)\\
        &=1-\sqrt{q}-\frac{\left(\eta ^2+\eta
   +1\right) q}{\eta }-\left(\eta
   +\frac{1}{\eta }+2\right)
   q^{3/2}-q^2+\frac{(\eta +1)^2
   \left(\eta ^2+1\right)
   q^{5/2}}{\eta ^2} \\
   &+\frac{2
   \left(\eta ^2+\eta +1\right)^2
   q^3}{\eta ^2}+\frac{\left(\eta
   ^4+5 \eta ^3+7 \eta ^2+5 \eta
   +1\right) q^{7/2}}{\eta
   ^2}-\frac{\left(\eta ^6+\eta ^5-4
   \eta ^4-6 \eta ^3-4 \eta ^2+\eta
   +1\right) q^4}{\eta ^3} \\
   &-\frac{2
   \left(\eta ^6+\eta ^5-\eta ^4-3
   \eta ^3-\eta ^2+\eta +1\right)
   q^{9/2}}{\eta
   ^3}+O\left(q^5\right).
    \end{aligned}
\end{equation}
For the round 3-sphere partition functions, using \eqref{eq:caldconptn}, we have($\simeq$ means equality up to an overall phase factor)
\begin{equation}
    \begin{aligned}
        \left(\mathcal{Z}^{\text{con}}_{b=1}\text{ of }\mathcal{T}_{SM(2,4r)}^{BG}\right)\simeq\left(\mathcal{Z}^{\text{con}}_{b=1}\text{ of }\frac{\mathcal{D}(4r,2r+1)}{\mathbb{Z}_{2}}\right)\simeq\sqrt{\frac{2}{r}}\sin\frac{\pi}{4r},
    \end{aligned}
\end{equation}
which again supports the proposed duality.

\subsubsection{$SM(3,6r-5)$}
For the case when $(P,Q)=(3,6r-5)$, the 3D $\mathcal{T}_{SM(P,Q)}$ is (we choose $R=-2,\,S=-4r+4$)
\begin{equation}
    \begin{aligned}
        &\mathcal{T}_{SM(3,6r-5)}\simeq T_{\text{irred}}[S^{2}((3,5),(6r-5,-4r+4),(3,1))]\simeq\mathcal{D}(6r-5,2r-1)\,.
    \end{aligned}
\end{equation}
Here we used \eqref{eq:dproperties} and the fact that $\mathcal{D}(3,\pm 1)$ are trivial theories. Using $\frac{2 r-1}{6r-5}=\frac{1}{3-\frac{1}{r-1-\frac{1}{-2}}}$, the $\mathcal{D}(6r-5,2r-1)$ is given as
\begin{equation}
    D(\vec{k})\simeq\mathcal{D}(6r-5,2r-1)\otimes\text{TFT}[\vec{k}]\text{ with }\vec{k}=(3,r-1,-2)\,.
\end{equation}
Meanwhile, in \cite{Baek:2025} an abelian $\mathcal{N}=2$ gauge theory description of $\mathcal{T}_{SM(3,6r-5)}$ was proposed(which we will call $\mathcal{T}_{SM(3,6r-5)}^{BG}$):
\begin{align}
\begin{split}
    &\mathcal{T}_{SM(3,6r-5)}^{BG}=\left(\frac{\mathcal{T}_\Delta^{\otimes r}}{[U(1)_\mathbf{Q}^r]_K}+ \text{monopole superpotentials}\right), \\
    &\text{with mixed CS level : } 
    K=
    2\left(\begin{array}{cccccc}
        \centering
             1&  -1&  -1& \ldots & -1&  -1  \\
             -1&  2&  2& \ldots & 2&  2  \\
             -1&  2&  4& \ldots & 4&  4  \\
             \vdots &  \vdots &  \vdots & \ddots & \vdots &  \vdots  \\
             -1&  2&  4& \ldots & 2(r-2)&  2(r-2) \\
             -1&  2&  4& \ldots & 2(r-2)&  2(r-1)
        \label{tab:my_label}
    \end{array}\right),\, \mathbf{Q}=\text{diag}(\bold{1}_{r-1},2).
\end{split}
\end{align}
Here $\mathcal{T}_{\Delta}$ is a free theory of a chiral multiplet with background CS level $-\frac{1}{2}$ for the $U(1)$ flavor symmetry. $\mathbf{Q}$ is a charge matrix such that $\mathbf{Q}_{ab}$ denotes the charge of $b$-th chiral multiplet under $a$-th $U(1)$ gauge symmetry.

We now claim that the two descriptions for $\mathcal{T}_{SM(3,6r-5)}$ are equivalent, namely
\begin{equation}
    \mathcal{T}_{SM(3,6r-5)}^{BG}\simeq\mathcal{D}(6r-5,2r-1)\,.
\end{equation}
We will check the equivalence by computing the various partition functions. First, the superconformal index for $D(\vec{k})$ can be computed using \eqref{eq:dkindex} and we find that
\paragraph{\boldmath \underline{$r=2$}} 
\begin{equation}
    \begin{aligned}
        &\mathcal{I}^{\text{sci}}_{D(\vec{k}=(3,1,-2))}(\eta,\nu=0)=\mathcal{I}_{\mathcal{T}^{BG}_{SM(3,7)}}^{\text{sci}}(q;\eta,\nu=0)\\
        &=1-q-\left(\eta+\frac{1}{\eta}\right)q^{3/2}-2q^{2}-\eta q^{5/2}+\left(\frac{1}{\eta^{2}}-1\right)q^{3}+\left(\frac{1}{\eta}-\eta\right)q^{7/2}+\frac{q^{4}}{\eta^{2}}\\
        &+\left(\eta+\frac{3}{\eta}\right)q^{9/2}+O\left(q^5\right),
    \end{aligned}
\end{equation}
\paragraph{\boldmath \underline{$r=3$}} 
\begin{equation}
    \begin{aligned}
        &\mathcal{I}^{\text{sci}}_{D(\vec{k}=(3,2,-2))}(\eta,\nu=0)=\mathcal{I}_{\mathcal{T}^{BG}_{SM(3,13)}}^{\text{sci}}(q;\eta,\nu=0)\\
        &=1-q-\left(\eta+\frac{1}{\eta}\right)q^{3/2}-q^{2}+\left(2\eta+\frac{2}{\eta}\right)q^{5/2}+\left(2\eta^{2}+\frac{2}{\eta^{2}}+6\right)q^{3}\\
        &+\left(6\eta+\frac{6}{\eta}\right)q^{7/2}+O\left(q^4\right),
    \end{aligned}
\end{equation}
\paragraph{\boldmath \underline{$r=4$}}
\begin{equation}
    \begin{aligned}
        &\mathcal{I}^{\text{sci}}_{D(\vec{k}=(3,3,-2))}(\eta,\nu=0)=\mathcal{I}_{\mathcal{T}^{BG}_{SM(3,19)}}^{\text{sci}}(q;\eta,\nu=0)\\
        &=1-q-\left(\eta+\frac{1}{\eta}\right)q^{3/2}-q^{2}+\left(2\eta+\frac{2}{\eta}\right)q^{5/2}+\left(2\eta^{2}+\frac{2}{\eta^{2}}+6\right)q^{3}+\left(4\eta+\frac{5}{\eta}\right)q^{7/2}\\
        &-\left(3\eta^{2}+\frac{1}{\eta^{2}}+1\right)q^{4}-\left(3\eta^{3}+7\eta+\frac{3}{\eta}+\frac{2}{\eta^{3}}\right)q^{9/2}+O\left(q^5\right).
    \end{aligned}
\end{equation}
For the round 3-sphere partition functions, using \eqref{eq:caldconptn}, we have($\simeq$ means equality up to an overall phase factor)
\begin{equation}
    \begin{aligned}
        \left(\mathcal{Z}^{\text{con}}_{b=1}\text{ of }\mathcal{T}_{SM(3,6r-5)}^{BG}\right)\simeq\left(\mathcal{Z}^{\text{con}}_{b=1}\text{ of }\mathcal{D}(6r-5,2r-1)\right)\simeq\frac{2}{\sqrt{6r-5}}\sin\left(\frac{\pi}{6r-5}\right)\,,
    \end{aligned}
\end{equation}
which again supports the proposed duality.
\subsubsection{$SM(3,6r-7)$}
For the case when $(P,Q)=(3,6r-7)$, the 3D $\mathcal{T}_{SM(P,Q)}$ is (we choose $R=-1,\,S=-2r+3$) 
\begin{equation}
    \begin{aligned}
        &\mathcal{T}_{SM(3,6r-7)}\simeq T_{\text{irred}}[S^{2}((3,4),(6r-7,-2r+3),(3,1))]\simeq\mathcal{D}(6r-7,-2r+3)\,.
    \end{aligned}
\end{equation}
Using $\frac{-2 r+3}{6r-7}=\frac{1}{-3-\frac{1}{r-1-\frac{1}{2}}}$, the $\mathcal{D}(6r-7,-2r+3)$ is given as
\begin{equation}
    D(\vec{k})\simeq\mathcal{D}(6r-7,-2r+3)\otimes\text{TFT}[\vec{k}]\text{ with }\vec{k}=(-3,r-1,2)\,.
\end{equation}
Meanwhile, in \cite{Baek:2025} an abelian $\mathcal{N}=2$ gauge theory description of $\mathcal{T}_{SM(3,6r-7)}$ was proposed(which we will call $\mathcal{T}_{SM(3,6r-7)}^{BG}$):
\begin{align}
\begin{split}
    &\mathcal{T}^{BG}_{SM(3,6r-7)}=\left(\frac{\mathcal{T}_\Delta^{\otimes r}}{[U(1)_\mathbf{Q}^r]_K}+ \text{monopole superpotentials}\right), \\
    &\text{with mixed CS level : } 
    K=
    2\left(\begin{array}{cccccc}
        \centering
             1&  -1&  -1& \ldots & -1&  -1  \\
             -1&  2&  2& \ldots & 2&  2  \\
             -1&  2&  4& \ldots & 4&  4  \\
             \vdots &  \vdots &  \vdots & \ddots & \vdots &  \vdots  \\
             -1&  2&  4& \ldots & 2(r-2)&  2(r-2) \\
             -1&  2&  4& \ldots & 2(r-2)&  2r
        \label{tab:my_label}
    \end{array}\right),\, \mathbf{Q}=\text{diag}(\bold{1}_{r-1},2).
\end{split}
\end{align}
We now claim that the two descriptions for $\mathcal{T}_{SM(3,6r-7)}$ are equivalent, namely
\begin{equation}
    \mathcal{T}^{BG}_{SM(3,6r-7)}\simeq\mathcal{D}(6r-7,-2r+3)\,.
\end{equation}
We will check the equivalence by computing the various partition functions. First, the superconformal index for $D(\vec{k})$ can be computed using \eqref{eq:dkindex} and we find that
\paragraph{\boldmath \underline{$r=3$}}
\begin{equation}
    \begin{aligned}
        &\mathcal{I}^{\text{sci}}_{D(\vec{k}=(-3,2,2))}(\eta,\nu=0)=\mathcal{I}_{\mathcal{T}^{BG}_{SM(3,11)}}^{\text{sci}}(q;\eta,\nu=0)\\
        &=1-q-\left(\eta+\frac{1}{\eta}\right)q^{3/2}-2q^{2}+\left(\eta^{2}+\frac{1}{\eta^{2}}+1\right)q^{3}+\left(2\eta+\frac{2}{\eta}\right)q^{7/2}\\
        &+\left(\frac{1}{\eta^{2}}+3\right)q^{4}+\left(-\eta^{3}+3\eta+\frac{5}{\eta}\right)q^{9/2}+O\left(q^5\right),
    \end{aligned}
\end{equation}
\paragraph{\boldmath \underline{$r=4$}} 
\begin{equation}
    \begin{aligned}
        &\mathcal{I}^{\text{sci}}_{D(\vec{k}=(-3,3,2))}(\eta,\nu=0)=\mathcal{I}_{\mathcal{T}^{BG}_{SM(3,17)}}^{\text{sci}}(q;\eta,\nu=0)\\
        &=1-q-\left(\eta+\frac{1}{\eta}\right)q^{3/2}-q^{2}+\left(2\eta+\frac{2}{\eta}\right)q^{5/2}+\left(2\eta^{2}+\frac{2}{\eta^{2}}+6\right)q^{3}\\
        &+\left(5\eta+\frac{6}{\eta}\right)q^{7/2}+O\left(q^4\right),
    \end{aligned}
\end{equation}
\paragraph{\boldmath \underline{$r=5$}} 
\begin{equation}
    \begin{aligned}
        &\mathcal{I}^{\text{sci}}_{D(\vec{k}=(-3,4,2))}(\eta,\nu=0)=\mathcal{I}_{\mathcal{T}^{BG}_{SM(3,23)}}^{\text{sci}}(q;\eta,\nu=0)\\
        &=1-q-\left(\eta+\frac{1}{\eta}\right)q^{3/2}-q^{2}+\left(2\eta+\frac{2}{\eta}\right)q^{5/2}+\left(2\eta^{2}+\frac{2}{\eta^{2}}+6\right)q^{3}\\
        &+\left(4\eta+\frac{5}{\eta}\right)q^{7/2}+O\left(q^4\right).
    \end{aligned}
\end{equation}
For the round 3-sphere partition functions, using \eqref{eq:caldconptn}, we have($\simeq$ means equality up to an overall phase factor)
\begin{equation}
    \begin{aligned}
        \left(\mathcal{Z}^{\text{con}}_{b=1}\text{ of }\mathcal{T}_{SM(3,6r-7)}^{BG}\right)\simeq\left(\mathcal{Z}^{\text{con}}_{b=1}\text{ of }\mathcal{D}(6r-7,-2r+3)\right)\simeq\frac{2}{\sqrt{6r-7}}\sin\left(\frac{\pi}{6r-7}\right)\,,
    \end{aligned}
\end{equation}
which again supports the proposed duality.

\section{Bulk field theory for $W_N$ minimal model $W_N(P,Q)$}
\label{sec:wn}
In this section, we begin by reviewing fundamental aspects of $W_{N}$ minimal models. Then, we propose the 3D bulk field theories for them, generalizing the results of the previous paper \cite{Gang:2024}.  

\subsection{$W_N$ minimal model $W_N(P,Q)$}

The 2D $W_N$ minimal model $W_N(P,Q)(=W_N(Q,P))$ is labeled by two integers $(P, Q)\in\mathds{Z}^2$ satisfying
\begin{align}\label{wn}
    P,Q\geq N, \text{ and }\, \gcd(P,Q)=1.
\end{align}
$W_N(P,Q)$ is unitary if $\abs{P-Q}=1$ and non-unitary otherwise. The $N=2$ case is the Virasoro minimal model.
 
\paragraph{\boldmath Primary operators}

Primary operators $\mathcal{O}_{\lambda\mu}$ of $W_N(P,Q)$ are labeled by the elements in $\Phi^{P,Q}_{W_N}$ defined as \cite{Beltaos:2012}:
\begin{align}
    \Phi^{P,Q}_{W_N}=\{\lambda\mu\in\Phi^{P}_{N}\times\Phi^{Q}_{N}: t(\mu)\equiv r\,t(\lambda)\,(\text{mod } N)\},
\end{align}
where 
\begin{align*}
&\Phi^{n}_{N}=\{\lambda=(\lambda_1,\cdots,\lambda_{N-1})\in\mathds{Z}^{N-1}: \lambda_i>0,\,\lambda_1+\cdots\lambda_{N-1}<n\}, \\
&r(\in\mathds{Z}): PQ-\text{admissible }r,\text{ chosen such that $rP-Q$ and $r^2P+Q$ are coprime to $2N$}, \\
&\text{and } t(\lambda):=\sum_{j=1}^{N-1}j\,\lambda_j: \text{$N$-ality}.
\end{align*}

\paragraph{Modular data} 

Modular data for the primary operators in $W_N(P,Q)$ are given as \cite{Beltaos:2012}:
\begin{align}
    S_{[\lambda\mu][\kappa\nu]}&=\text{(const.)}\,\text{exp}\left(-\frac{2\pi i}{N}(t(\lambda)t(\nu)+t(\mu)t(\kappa))\right)S^{(N;P/Q)}_{\lambda\kappa}S^{(N;Q/P)}_{\mu\nu}, \\
    T_{[\lambda\mu][\lambda\mu]}&=\text{(const.)}\,\text{exp}\left(\frac{\pi i}{PQ}(Q\lambda-P\mu)\cdot(Q\lambda-P\mu)\right),
\end{align}
where
\begin{align*}
&S_{\lambda\mu}^{(N;n)}:=\text{(const.)}\,\text{exp}\left(\frac{2\pi i}{Nn}t(\lambda)t(\mu)\right)\text{det}_{1\leq i,j\leq N}(\text{exp}\left(-\frac{2\pi i}{n}\lambda[i]\mu[j]\right)), \\
&\lambda[i]:=\sum_{i\leq l<N}(\lambda_l+1), \\
&\text{and } \lambda\cdot\mu:=\sum_{1\leq i<N}\frac{i(N-i)}{N}\lambda_i\mu_i+\sum_{1\leq i<j<N}\frac{i(N-j)}{N}(\lambda_i\mu_j+\lambda_j\mu_i).
\end{align*} 

\subsection{Bulk field theory as $T_{N,\text{irred}}[M]$ with $M=S^2((P,P-R),(Q,S),(N+1,-2N-1))$}
As in section \ref{sec:smbulkdual}, we will define $\mathcal{T}_{W_{N}(P,Q)}$ as a bulk theory such that $R[\mathcal{T}_{W_{N}(P,Q)},\mathbb{B}]=W_{N}(P,Q)$ for some proper $\mathbb{B}$.
\begin{equation}
    \begin{aligned}
        &\textbf{Def: 3D }\mathcal{T}_{W_{N}(P,Q)}\textbf{ theory is defined as}\\
        &\text{For }|P-Q|=1,\,\mathcal{T}_{W_{N}(P,Q)}\text{ is a 3D unitary TQFT with}\\
        &\mathcal{T}_{W_{N}(P,Q)}\xrightarrow{\text{at boundary with a proper }\mathbb{B}}W_{N}(P,Q),\\
        &\text{For }|P-Q|>1,\,\mathcal{T}_{W_{N}(P,Q)}\text{ is a }\mathcal{N}=4\text{ 3D rank-0 SCFT with}\\
        &\mathcal{T}_{W_{N}(P,Q)}\xrightarrow{\text{a top'l twisting}}\text{non-unitary TQFT }\mathcal{T}^{\text{top}}_{W_{N}(P,Q)}\xrightarrow{\text{at bdy. with a proper }\mathbb{B}}W_{N}(P,Q)\,.
    \end{aligned}
\end{equation}
Note that $W_{N}(P, Q)$ is a bosonic CFT. Basic dictionaries of the bulk-boundary correspondence are given in the first and second columns of Table~\ref{tab:wndict}.

We will obtain $\mathcal{T}_{W_{N}(P,Q)}$ theory via 3D-3D correspondence. In particular, we will use the $T_{N,\text{irred}}[M]$ theory with $M=S^{2}(\vec{p},\vec{q})$. The theory is believed to describe an effective 3D field theory of 6D $A_{N}\,\mathcal{N}=(2,0)$ superconformal field theory compactified on the 3-manifold $M$(in Section~\ref{sec:sm} we set $N=2$). As we mentioned in Section~\ref{sec:smbulkdual}, $T_{N,\text{irred}}[M]$ only sees the $SL(N,\mathbb{C})$ irreducible flat connections on $M$. 
\begin{table}
\centering
\begin{tabular}{ |c|c|c| } 
 \hline
 2D chiral RCFT $\chi\mathcal{R}[M;\mathbb{B}]$ & $T_{N,\text{irred}}[M]$ & $SL(N,\mathbb{C})$ CS on $M$ \\ 
 \hline
 (unitary)/(non-unitary) & (mass gap)/(rank-0 SCFT) & equation~\eqref{eq:irphase}  \\ 
 \hline
 Primary $\mathcal{O}_{\alpha=0,\dots,n-1}$ & Bethe-vacuum $\vec{x}_{\alpha}\in\mathcal{S}_{\text{BE}}$ & $\rho_{\alpha}\in\chi_{N,\text{irred}}[M]$\\
 \hline
 Conformal dimension $h_{\alpha}$ & $e^{2\pi ih_{\alpha}}=\mathcal{F}^{\text{top}}(\vec{x}_{\alpha})/\mathcal{F}^{\text{top}}(\vec{x}_{\alpha=0})$ & $e^{2\pi ih_{\alpha}}=e^{2\pi i(\text{CS}[\rho_{\alpha=0}]-\text{CS}[\rho_{\alpha}])}$\\
 \hline
 $S_{0\alpha}^2$& $(\mathcal{H}^{\text{top}}(\vec{x}_{\alpha}))^{-1}$&$1/(2\text{Tor}[\rho_{\alpha}])$\\
 \hline
 $|S_{00}|$&$|\mathcal{Z}_{b}^{\text{top}}|$&$|\sum_{\rho_{\alpha}}\frac{e^{-2\pi i\text{CS}[\rho_{\alpha}]}}{2\text{Tor}[\rho_{\alpha}]}|$\\
 \hline
 $\min_{\alpha}\{|S_{0\alpha}|\}$&$e^{-F}=|\mathcal{Z}_{b=1}^{\text{con}}|$&$\min_{\alpha}\{1/\sqrt{2\text{Tor}[\rho_{\alpha}]}\}$\\
 \hline
\end{tabular}
\caption{Basic dictionaries (adopted from \cite{Gang:2024}) for the correspondence among topological data of non-hyperbolic 3-manifolds $M$, BPS partition functions of 3D bulk theories $T_{N,\text{irred}}[M]$, and conformal data of 2D chiral RCFTs $\chi\mathcal{R}[M]$ for $M=S^{2}(\vec{p},\vec{q})$ with trivial $H^{1}(M,\mathbb{Z}_{2})$. For the notations see the caption of Table~\ref{tab:ferdict}.}
\label{tab:wndict}
\end{table}
Basic dictionaries of the 3D-3D correspondence are given in Table~\ref{tab:wndict}. Refer to \cite{Gang:2019,Gang:2019uay,Benini:2019dyp,Cho:2020ljj,Cui:2021lyi,Bonetti:2024} for details. Note that, as in Section~\ref{sec:smbulkdual} the $SL(N,\mathbb{C})$ flat connections on $M$ can be alternatively described by a homomorphism $\rho\in\chi_{N,\text{irred}}[M]$. CS[$\rho$] is defined mod 1. For $M=S^{2}(\vec{p},\vec{q})$ with trivial $H^{1}(M,\mathbb{Z}_{2})$, one can determine the IR phase of the $T_{N,\text{irred}}[M]$ theory in the following way \cite{Choi:2022dju,Cho:2020ljj}
\begin{equation}
\label{eq:irphase}
    \begin{aligned}
        &T_{N,\text{irred}}[M=S^{2}(\vec{p},\vec{q})]\\
        &\xrightarrow{\text{in the IR}}\begin{cases}
            \text{unitary TQFT},& \text{if }|\sum_{\rho\in\chi}\frac{e^{-2\pi iCS[\rho_{\alpha}]}}{2\text{Tor}[\rho]}|\leq\frac{1}{\sqrt{|2\text{Tor}[\rho_{\alpha}]|}}\quad\forall\rho_{\alpha}\in\chi,\\
            \text{3D rank-0 SCFT,}& \text{otherwise.}
        \end{cases}
    \end{aligned}
\end{equation}
Here, $\chi$ abbreviates $\chi_{N,\text{irred}}[M]$.

Our main conjecture in this section is given as follows:
\begin{align} \label{eq:shift}
    \mathcal{T}_{W_N(P,Q)}\simeq T_{N,\text{irred}}[S^2((P,P-R),(Q,S),(N+1,-2N-1))]\,,
\end{align}
where $PS-QR=1$ and $\gcd(P,R)=\gcd(Q,S)=1$. It fixes the $(R,S)$ modulo a shift $(R,S)\rightarrow(R,S)+\mathbb{Z}(P,Q)$. Though we won't provide a rigorous proof, we expect that \eqref{eq:shift} will hold for a general value of $N$.

Let us check the proposal using the dictionaries in Table~\ref{tab:wndict}. Unlike the $SL(2,\mathbb{C})$ case, it is generally hard to compute $SL(N,\mathbb{C})$ flat connections for $N\geq 3$. Despite that, we can still compute some of them by embedding the $SL(2,\mathbb{C})$ flat connections. First, note that the $SL(2,\mathbb{C})$ connections can be written as
\begin{equation}
    \mathcal{A}=\sum_{a=1}^{3}\mathcal{A}^{a}t^{a},\,t^{a}=\frac{1}{2}\sigma^{a}\,,
\end{equation}
where $\sigma^{a}$ are Pauli matrices. But this can also be written in the $N$-dimensional irreducible representation. Denoting the generators in $N$-dimensional irreducible representation as $\mathcal{R}(t^{a})$, we get 
\begin{equation}
\mathcal{A}^{SL(N,\mathbb{C})}=\sum_{a=1}^{3}\mathcal{A}^{a}\mathcal{R}(t^{a})\,.
\end{equation}
Recall that
\begin{equation}
    \text{CS}[\rho]\coloneq\frac{1}{8\pi^{2}}\int\Tr\left(\mathcal{A}_{\rho}d\mathcal{A}_{\rho}+\frac{2}{3}\mathcal{A}_{\rho}^{3}\right)\pmod{1}\,.
\end{equation}
CS invariant of $\mathcal{A}^{SL(N,\mathbb{C})}$ can be computed as
\begin{equation}
\begin{aligned}
\label{eq:slncsinv}
    \text{CS}[\rho ,SL(N,\mathbb{C})]&=\frac{\Tr[\mathcal{R}(t^{a})\mathcal{R}(t^{b})]}{\Tr [t^{a}t^{b}]}\text{CS}[\rho ,SL(2,\mathbb{C})]\\
    &=\frac{N(N^{2}-1)}{6}\text{CS}[\rho ,SL(2,\mathbb{C})]\,,
    \end{aligned}
\end{equation}
where we used the fact that
\begin{equation}
    \Tr[\mathcal{R}(t^{a})\mathcal{R}(t^{b})]=\frac{N(N^{2}-1)}{12}\delta^{ab}\,.
\end{equation}
But not all of these lifted $SL(N,\mathbb{C})$ connections are acceptable. First, in addition to being irreducible, the connection should be `anyonic', in the sense that for each $k=1,2,3$, the eigenvalues of the matrix $\rho (x_{k})$ are all distinct  \cite{Bonetti:2024}. Otherwise, the solution transforms into itself under the Weyl symmetry transformation, and does not belong to the Bethe vacua. Secondly, the connections may become equivalent after embedding, even if we started with the inequivalent $SL(2,\mathbb{C})$ connections.

For example, consider $W_{3}(3,7)$. Setting $(P,Q,R,S)=(3,7,-1,-2)$, the corresponding Seifert manifold is $S^{2}((3,4),(7,-2),(4,-7))$. There are nine $SL(2,\mathbb{C})$ irreducible flat connections, whose CS invariants are $\left( \frac{251}{336},\frac{299}{336},\frac{59}{336},\frac{83}{336},\frac{131}{336},\frac{227}{336},\frac{83}{84},\frac{59}{84},\frac{47}{84}\right)$. After embedding, the last three are not anyonic, while the first three and middle three are equivalent. Thus, we get three distinct $SL(3,\mathbb{C})$ irreducible flat connections with CS invariant $\left( \frac{83}{84},\frac{47}{84},\frac{59}{84}\right)$. In fact, $W_{3}(3,7)$ has five primaries with spin $(0,-\frac{3}{7},\frac{2}{7},\frac{3}{7},-\frac{3}{7})$. Two of these cannot be obtained from the embedding. 

In Tables \ref{tab:w337} - \ref{tab:w447}, we show independent, irreducible, and anyonic embedded connections and their CS invariants. Unfortunately, we could not figure out the one-to-one map between embedded connections and primary operators. Despite that, to confirm whether embedded connections are compatible with the spins of primary operators, we assumed that $\rho_{\vec{j}=0}$ corresponds to the identity operator and computed $\text{CS}_{N}[\rho_{\vec{j}=0}]-\text{CS}_{N}[\rho]$. In all five cases, they agreed well.

\begin{table}
\centering
\begin{tabular}{ |c|c|c|c| } 
 \hline
 $(j_{1},j_{2},j_{3})$ & $(\vec{n},\lambda)$ & $\text{CS}_{N}[\rho]\pmod{1}$ & $\text{CS}_{N}[\rho_{\vec{j}=0}]-\text{CS}_{N}[\rho]\pmod{1}$ \\ 
 \hline
 $(0,0,0)$ & $(1,3,\frac{1}{2},\frac{1}{2})$ & $\frac{83}{84}$ & 0 \\ 
 \hline
 $(0,2,0)$ & $(1,2,\frac{1}{2},\frac{1}{2})$ & $\frac{47}{84}$ & $\frac{3}{7}$ \\
 \hline
 $(0,4,0)$ & $(1,1,\frac{1}{2},\frac{1}{2})$ & $\frac{59}{84}$ & $\frac{2}{7}$ \\
 \hline
\end{tabular}
\caption{$W_{3}(3,7)$ from $S^{2}((3,4),(7,-2),(4,-7))$. $(j_{1},j_{2},j_{3})$ and $(\vec{n},\lambda)$ characterizes the $SL(2,\mathbb{C})$ connections. For more details, see \cite{Gang:2024}. We assumed that $\vec{j}=0$ corresponds to the vacuum. The result is compatible with that $h=0,\frac{4}{7},\frac{2}{7},\frac{3}{7},\frac{4}{7}\pmod{1}\,.$}
\label{tab:w337}
\end{table}

\begin{table}
\centering
\begin{tabular}{ |c|c|c|c| } 
 \hline
 $(j_{1},j_{2},j_{3})$ & $(\vec{n},\lambda)$ & $\text{CS}_{N}[\rho]\pmod{1}$ & $\text{CS}_{N}[\rho_{\vec{j}=0}]-\text{CS}_{N}[\rho]\pmod{1}$ \\ 
 \hline
 $(0,0,0)$ & $(1,\frac{1}{2},\frac{1}{2},\frac{1}{2})$ & $\frac{17}{24}$ & 0 \\ 
 \hline
 $(0,2,0)$ & $(1,\frac{3}{2},\frac{1}{2},\frac{1}{2})$ & $\frac{17}{24}$ & $0$ \\
 \hline
\end{tabular}
\caption{$W_{3}(3,8)$ from $S^{2}((3,2),(8,3),(4,-7))$. We assumed that $\vec{j}=0$ corresponds to the vacuum. The result is compatible with that $h=0,\frac{1}{4},\frac{1}{8},\frac{1}{2},0,\frac{1}{4},\frac{1}{2}\pmod{1}\,.$}
\label{tab:w338}
\end{table}

\begin{table}
\centering
\begin{tabular}{ |c|c|c|c| } 
 \hline
 $(j_{1},j_{2},j_{3})$ & $(\vec{n},\lambda)$ & $\text{CS}_{N}[\rho]\pmod{1}$ & $\text{CS}_{N}[\rho_{\vec{j}=0}]-\text{CS}_{N}[\rho]\pmod{1}$ \\ 
 \hline
 $(0,0,0)$ & $(\frac{1}{2},\frac{1}{2},\frac{1}{2},\frac{1}{2})$ & $\frac{7}{10}$ & 0 \\ 
 \hline
 $(0,2,0)$ & $(\frac{1}{2},\frac{3}{2},\frac{1}{2},\frac{1}{2})$ & $\frac{3}{10}$ & $\frac{2}{5}$ \\
 \hline
\end{tabular}
\caption{$W_{3}(4,5)$ from $S^{2}((4,5),(5,-1),(4,-7))$. We assumed that $\vec{j}=0$ corresponds to the vacuum. The result is compatible with that $h=0,\frac{2}{5},\frac{2}{3},\frac{1}{15},\frac{2}{3},\frac{1}{15}\pmod{1}\,.$}
\end{table}

\begin{table}
\centering
\begin{tabular}{ |c|c|c|c| } 
 \hline
 $(j_{1},j_{2},j_{3})$ & $(\vec{n},\lambda)$ & $\text{CS}_{N}[\rho]\pmod{1}$ & $\text{CS}_{N}[\rho_{\vec{j}=0}]-\text{CS}_{N}[\rho]\pmod{1}$ \\ 
 \hline
 $(0,0,0)$ & $(\frac{1}{2},3,\frac{1}{2},\frac{1}{2})$ & $\frac{3}{7}$ & 0 \\ 
 \hline
 $(0,2,0)$ & $(\frac{1}{2},2,\frac{1}{2},\frac{1}{2})$ & $\frac{6}{7}$ & $\frac{4}{7}$ \\
 \hline
 $(0,4,0)$ & $(\frac{1}{2},1,\frac{1}{2},\frac{1}{2})$ & $\frac{5}{7}$ & $\frac{5}{7}$ \\
 \hline
\end{tabular}
\caption{$W_{3}(4,7)$ from $S^{2}((4,3),(7,2),(4,-7))$. We assumed that $\vec{j}=0$ corresponds to the vacuum. The result is compatible with that $h=0,\frac{3}{7},\frac{5}{7},\frac{4}{7},\frac{3}{7},\frac{1}{3},\frac{16}{21},\frac{1}{21},\cdots\pmod{1}\,.$}
\label{tab:w347}
\end{table}

\begin{table}
\centering
\begin{tabular}{ |c|c|c|c| } 
 \hline
 $(j_{1},j_{2},j_{3})$ & $(\vec{n},\lambda)$ & $\text{CS}_{N}[\rho]\pmod{1}$ & $\text{CS}_{N}[\rho_{\vec{j}=0}]-\text{CS}_{N}[\rho]\pmod{1}$ \\ 
 \hline
 $(0,0,0)$ & $(\frac{1}{2},3,\frac{1}{2},\frac{1}{2})$ & $\frac{11}{56}$ & 0 \\ 
 \hline
 $(0,2,0)$ & $(\frac{1}{2},2,\frac{1}{2},\frac{1}{2})$ & $\frac{43}{56}$ & $\frac{3}{7}$ \\
 \hline
 $(0,4,0)$ & $(\frac{1}{2},1,\frac{1}{2},\frac{1}{2})$ & $\frac{51}{56}$ & $\frac{2}{7}$ \\
 \hline
\end{tabular}
\caption{$W_{4}(4,7)$ from $S^{2}((4,3),(7,2),(5,-9))$. We assumed that $\vec{j}=0$ corresponds to the vacuum. The result is compatible with that $h=0,\frac{4}{7},\frac{3}{7},\frac{2}{7},\frac{4}{7}\pmod{1}\,.$}
\label{tab:w447}
\end{table}

\subsection{Field theory description of $\mathcal{T}_{W_N(P,Q)}$}
Construction of $T_{N}[S^{2}(\vec{p},\vec{q})]$ is studied in \cite{Assel:2022row}. In the previous paper  \cite{Gang:2024}, they found the concrete field theory description of $T_{N=2,\text{irred}}[S^{2}(\vec{p},\vec{q})]$. Here, we extend their proposal and conjecture the field theory description for $T_{N,\text{irred}}[S^{2}(\vec{p},\vec{q})]$ for general $N$. Though the construction in  \cite{Gang:2024} can be straightforwardly generalized to the general $N$ case, we repeat it here for completeness.

\begin{figure}[htbp]
\centering
\includegraphics[width=\textwidth]{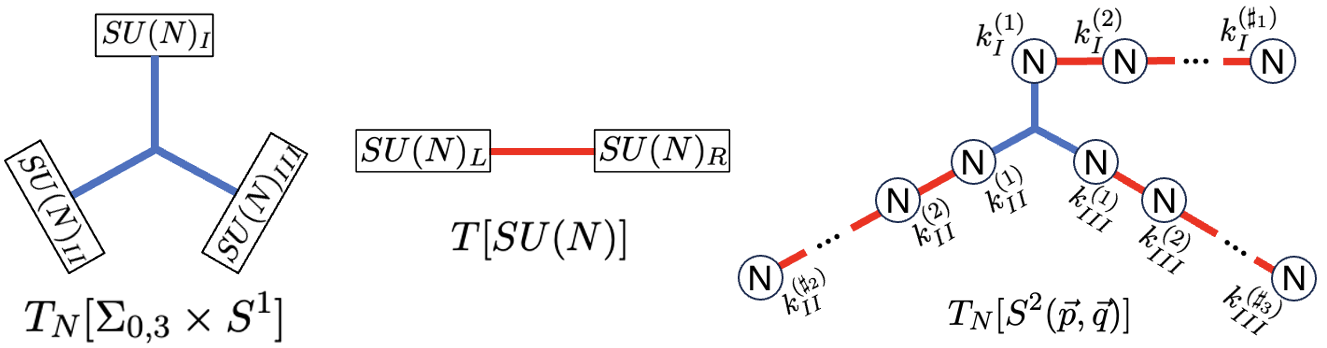}
\caption{Generalized quiver diagrams for $T_{N}[\Sigma_{0,3}\times S^{1}],T[SU(N)]$ and $T_{N}[S^{2}(\vec{p},\vec{q})]$. The difference between $T_{N,\text{full}}[S^{2}(\vec{p},\vec{q})]$ and $T_{N,\text{irred}}[S^{2}(\vec{p},\vec{q})]$ arises from different choices of the $T_{N}[\Sigma_{0,3}\times S^{1}]$ theory, either $T_{N,\text{full}}[\Sigma_{0,3}\times S^{1}]$ or $T_{N,\text{irred}}[\Sigma_{0,3}\times S^{1}]$.}
\label{fig:tn}
\end{figure}
$T_{N}[\Sigma_{0,3}\times S^{1}]$ is expected to be a 3d $\mathcal{N}=4$ theory with three $SU(N)$ flavor symmetries associated with three boundary tori. In Fig.~\ref{fig:tn}, the theory is depicted by a blue trivalent vertex with three legs, and the boxes attached to the legs represent the three $SU(N)$ flavor symmetries. Consider gluing the solid torus by the Dehn filling procedure. Denote the Dehn filling slope by $(p,q)$ and corresponding CS levels $\vec{k}=(k^{(1)},\cdots,k^{(\sharp)})$. They are related by the equation \eqref{eq:dehnslope}. Corresponding field-theoretic operation is done by gluing $\sharp-1$ $T[SU(N)]$ theories. Here, $T[SU(N)]$ is a 3d $\mathcal{N}=4$ theory with $SU(N)_{L}\times SU(N)_{R}$ flavor symmetry. When gluing, we diagonally gauge the $2\sharp -1$ $SU(N)$ symmetries(one from $T[\Sigma_{0,3}\times S^{1}]$, $2(\sharp-1)$ from $T[SU(N)]$ theories) in the manner described in Fig.~\ref{fig:tn}. The circle denotes the $\mathcal{N}=3$ gauging of the diagonal $SU(N)$, and the integer $k$ next to the circle denotes the CS level.

The above construction was studied in  \cite{Assel:2022row}, in the context of $T_{\text{full}}$ theory. In particular, they start with $T_{\text{full}}[\Sigma_{0,3}\times S^{1}]$, 4d $T_{N}$ theory compactified on $S^{1}$ which has $SU(N)^{3}$ flavor symmetry. But we are interested in $T_{\text{irred}}$ theory, which is the low-energy limit of $T_{\text{full}}$. For the difference between $T_{\text{full}}$ and $T_{\text{irred}}$ refer to  \cite{Gang:2018wek}. $T_{\text{irred}}$ can be constructed in the same way as above, starting with an appropriate $T_{\text{irred}}[\Sigma_{0,3}\times S^{1}]$ theory.

In  \cite{Gang:2024}, the field theory description for the $N=2$ case was proposed up to the stronger IR equivalence(see \eqref{eq:tirred} and \eqref{eq:calddef}). However, for general $N$, we could not determine the decoupled TQFT. Thus, for general $N$ we propose the field theory description only up to the `topological operations($\sim$)', which include tensoring with a unitary TQFT or gauging of the finite symmetries.
\begin{equation}
\label{eq:wnproposal}
    \begin{aligned}
        \mathcal{T}_{W_N(P,Q)}\simeq T_{N,\text{irred}}[S^2((P,P-R),(Q,S),(N+1,-2N-1))]\sim D_{N}(\vec{k_{1}})\otimes D_{N}(\vec{k_{2}})\otimes D_{N}(\vec{k_{3}})\,.
    \end{aligned}
\end{equation}
Here $\sim$ denotes the weak IR equivalence. The theory $D_{N}(\vec{k})$ is defined as follows:
\begin{equation}
    \begin{aligned}
        D_{N}(\vec{k})\coloneq\begin{cases}
            \frac{T[SU(N)]^{\otimes(\sharp-1)}}{SU(N)^{(1)}_{k^{(1)}}\times SU(N)^{(2)}_{k^{(2)}}\dots\times SU(N)^{(\sharp)}_{k^{(\sharp)}}}, & \sharp\geq 2\\
            \mathcal{N}=2\text{ pure }SU(N)_{k^{(1)}}\text{ CS theory}, & \sharp=1
            \end{cases}\,.
    \end{aligned}
\end{equation}
Here $/G_{k}$ denotes $\mathcal{N}=3$ gauging of $G$ symmetry with CS level $k$. Recall that $\vec{k}$ and $(p,q)$ are related by \eqref{eq:dehnslope}. The gauged $SU(N)$ symmetries are
\begin{equation}
\begin{aligned}
    &SU(N)^{(1)}:SU(N)_{L}^{(1)}\coloneq SU(N)_{L}\text{ of the 1st }T[SU(N)]\,,\\
    &SU(N)^{(2\leq I\leq\sharp -1)}:\text{diagonal subgroup of }(SU(N)_{R}^{(I-1)}\times SU(N)_{L}^{(I)})\,,\\
    &SU(N)^{(\sharp)}:SU(N)_{R}^{(\sharp -1)}\,.
\end{aligned}
\end{equation}
Let us check the relation \eqref{eq:wnproposal}. Unfortunately, it is difficult to compute the superconformal index. Instead, we will compute the $\mathcal{Z}^{\text{con}}_{b=1}$ and compare them with $\min|S_{0\alpha}|$. From \eqref{eq:n3zcon}, for $N=3$, $\mathcal{Z}^{\text{con}}_{b=1}$ of the $D_{N}(\vec{k_{1}})\otimes D_{N}(\vec{k_{2}})\otimes D_{N}(\vec{k_{3}})$ theory is
\begin{equation}
\label{eq:d3kzcon}
    \mathcal{Z}^{\text{con}}_{b=1}=(\text{const.})\times\frac{1}{PQ}\sin^{2}\left(\frac{\pi}{P}\right)\sin\left(\frac{2\pi}{P}\right)\sin^{2}\left(\frac{\pi}{Q}\right)\sin\left(\frac{2\pi}{Q}\right)\,.
\end{equation}
Meanwhile, $|S_{0\alpha}|$ of the $W_{3}(P,Q)$ minimal model is \cite{Beltaos:2012}
\begin{equation}
\begin{aligned}
    &|S_{[\lambda\mu][\rho\rho]}|=|\frac{\sqrt{3}}{PQ}S_{\lambda\rho}^{(P/Q)}S_{\mu\rho}^{(Q/P)}|\\
    &\text{where }S^{(n)}_{\lambda\rho}=\frac{8}{\sqrt{3}n}\sin\left(\frac{\pi\lambda_{1}}{n}\right)\sin\left(\frac{\pi\lambda_{2}}{n}\right)\sin\left(\frac{\pi(\lambda_{1}+\lambda_{2})}{n}\right)\,.
    \end{aligned}
\end{equation}
The overall constant of the first equation is fixed experimentally. Accordingly,
\begin{equation}
\label{eq:min}
    \min|S_{[\lambda\mu][\rho\rho]}|=\frac{64}{\sqrt{3}PQ}\sin^{2}\left(\frac{\pi}{P}\right)\sin\left(\frac{2\pi}{P}\right)\sin^{2}\left(\frac{\pi}{Q}\right)\sin\left(\frac{2\pi}{Q}\right)\,.
\end{equation}
\eqref{eq:min} is consistent with \eqref{eq:d3kzcon} up to the overall constant.

\section{Conclusion and Future Directions}
In this paper, we proposed 3D $\mathcal{T}_{SM(P, Q)}$ theory corresponding to the $\mathcal{N}=1$ supersymmetric minimal model $SM(P, Q)$, along with the field theory descriptions. We also conjectured the 3D bulk theories for $W_{N}$ minimal models. The main results are presented in \eqref{eq:fielddescription} and \eqref{eq:wnproposal}. We suggest some future directions below.
\paragraph{Further consistency checks} For $SM(P,Q)$ theory, we couldn't say much about the $\Tilde{\mathcal{T}}_{SM(P,Q)}$ theory(bosonic mother theory of $\mathcal{T}_{SM(P,Q)}$). If we could compute the CS invariant and torsion of the $PSL(2,\mathbb{C})$ flat connections, it would be possible to analyze the bosonic theories. Additionally, computing the half-indices of the bulk field theory and comparing them with the conformal characters will provide strong evidence for the duality.  
\paragraph{Generalization to various Seifert manifolds} Consider the $T_{N,\text{irred}}[M]$ theory for the manifold $S^{2}((P,P-R),(Q,S),(N+1,-2N-1))$. We now know that $N=2, PS-QR=1$ corresponds to the Virasoro minimal model, and $N=2, PS-QR=2$ gives the $\mathcal{N}=1$ supersymmetric minimal model. Generalization to arbitrary $PS-QR$ value seems promising; they may give another class of rational CFTs, such as parafermion CFTs. Also, it would be interesting to see what happens when we set both $N$ and $PS-QR$ to arbitrary values.
\acknowledgments
We thank Dongmin Gang for initial collaboration and numerous valuable comments. The work of SB and HK is supported in part by the National Research Foundation of Korea (NRF) under grants NRF-2022R1C1C1011979 and RS-2024-00405629.
\appendix
\section{BPS partition functions of $D(\vec{k})$}
In Appendix A, we recapitulate some material from \cite{Gang:2024,Gang:2021hrd} for completeness. We recall that the $T[SU(2)]$ theory is a 3D $\mathcal{N}=4$ SQED with two fundamental hypermultiplets. In terms of the $\mathcal{N}=2$ algebra, it contains five chiral multiplets---two from each $\mathcal{N}=4$ hypermultiplet and an adjoint chiral multiplet from the $\mathcal{N}=4$ vector multiplet. The quantum numbers of the $\mathcal{N}=2$ chiral multiplets are listed in Table~\ref{tab:tsu2multiplets}. Partition functions of the $T[SU(2)]$ theory have been computed in various literature; we briefly review the relevant results.
\begin{table}
\centering
\begin{tabular}{ |c|c|c|c|c| } 
 \hline
 Chiral multiplet & $U(1)_{\text{gauge}}$ & $R_{\nu=0}$ & $A$&$F$ \\ 
 \hline
 $(\Phi_{1},\Phi_{2})$ & $(+1,-1)$ & $\frac{1}{2}$ & $\frac{1}{2}$&$(+1,-1)$ \\ 
 \hline
 $(\Phi_{3},\Phi_{4})$ & $(+1,-1)$ & $\frac{1}{2}$ & $\frac{1}{2}$&$(-1,+1)$ \\ 
 \hline
 $\Phi_{0}$ & $0$ & $1$ & $-1$ &$0$\\
 \hline
\end{tabular}
\caption{$\mathcal{N}=2$ chiral multiplets in the $T[SU(2)]$ theory. $(\Phi_{1},\Phi_{2})$ and $(\Phi_{3},\Phi_{4})$ each constitute $\mathcal{N}=4$ hypermultiplet, and $\Phi_{0}$ is from the $\mathcal{N}=4$ vector multiplet. $(\Phi_{1},\Phi_{2})$ and $(\Phi_{3},\Phi_{4})$ transform as $\mathbf{2}$ under $SU(2)_{L}$ flavor symmetry. $F$ is a Cartan of $SU(2)_{L}$. $A$ is a topological symmetry associated to $U(1)_{\text{gauge}}$, which enhances to the $SU(2)_{R}$ flavor symmetry in the IR. $R_{\nu=0}$ is the R-symmetry of the $\mathcal{N}=2$ algebra. $T[SU(2)]$ theory also has a superpotential $\mathcal{W}\propto\Phi_{1}\Phi_{0}\Phi_{2}-\Phi_{3}\Phi_{0}\Phi_{4}$.}
\label{tab:tsu2multiplets}
\end{table}
\paragraph{Superconformal index  \cite{Kim:2009wb,Imamura:2011su,Dimofte:2011py}}
The generalized superconformal index $\mathcal{I}_{T[SU(2)]}^{\text{sci}}$ for the $T[SU(2)]$ theory is $(\tilde{\eta}^{2}\coloneq\eta)$
\begin{equation}
\label{eq:tsu2index}
\begin{aligned}
    &\mathcal{I}_{T[SU(2)]}^{\text{sci}}(a_{1},a_{2},\eta,\nu;m_{1},m_{2})\\
    &=\sum_{n}\oint_{|u|=1}\frac{du}{2\pi iu}((-1)^{m_{2}}a_{2})^{-2n}((-1)^{n}u)^{-2m_{2}}(-q^{\frac{1}{2}}\eta^{-1})^{\frac{1}{2}(|m_{1}+n|+|m_{1}-n|)}\\
    &\times\frac{((-q^{\frac{1}{2}})^{\frac{3}{2}+|m_{1}+n|}\eta^{-\frac{1}{2}}a_{1}u;q)_{\infty}((-q^{\frac{1}{2}})^{\frac{3}{2}+|m_{1}+n|}\eta^{-\frac{1}{2}}a_{1}^{-1}u^{-1};q)_{\infty}}{((-q^{\frac{1}{2}})^{\frac{1}{2}+|m_{1}+n|}\eta^{\frac{1}{2}}a_{1}u;q)_{\infty}((-q^{\frac{1}{2}})^{\frac{1}{2}+|m_{1}+n|}\eta^{\frac{1}{2}}a_{1}^{-1}u^{-1};q)_{\infty}}\\
    &\times\frac{((-q^{\frac{1}{2}})^{\frac{3}{2}+|m_{1}-n|}\eta^{-\frac{1}{2}}a_{1}u^{-1};q)_{\infty}((-q^{\frac{1}{2}})^{\frac{3}{2}+|m_{1}-n|}\eta^{-\frac{1}{2}}a_{1}^{-1}u;q)_{\infty}(-q^{\frac{1}{2}}\eta;q)_{\infty}}{((-q^{\frac{1}{2}})^{\frac{1}{2}+|m_{1}-n|}\eta^{\frac{1}{2}}a_{1}u^{-1};q)_{\infty}((-q^{\frac{1}{2}})^{\frac{1}{2}+|m_{1}-n|}\eta^{\frac{1}{2}}a_{1}^{-1}u;q)_{\infty}(-q^{\frac{1}{2}}\eta^{-1};q)_{\infty}}\Bigg\rvert_{\eta\rightarrow\eta(-q^{\frac{1}{2}})^{\nu}}\,.
\end{aligned}
\end{equation}
In the last line, we changed the integration variable $u$ to $u(-q^{1/2})^{1/2}$. This amounts to mixing the $U(1)$ R-symmetry with the $U(1)$ gauge symmetry:
\begin{equation}
\label{eq:gaugemix}
    R_{\nu}\rightarrow\tilde{R_{\nu}}\coloneq R_{\nu}+\frac{1}{2}G\,.
\end{equation}
\eqref{eq:gaugemix} does not change the result because the index counts the gauge invariant operators. Change of variable makes \eqref{eq:tsu2index} easier to handle using Mathematica. The index depends on the following parameters:
\begin{equation}
    \begin{aligned}
        &(m_{1},u_{1})/(m_{2},u_{2})\colon\text{(monopole flux, fugacity) for the Cartans of }SU(2)_{L}/SU(2)_{R}\,,\\
        &(\eta,\nu)\colon\text{fugacity and R-symmetry mixing parameter for the }U(1)_{A}\text{ symmetry}\,.
    \end{aligned}
\end{equation}
$\mathcal{I}_{\Delta}(m,u)$ denotes the tetrahedron index:
\begin{equation}
    \begin{aligned}
        &\mathcal{I}_{\Delta}(m,u)\coloneq\prod_{r=0}^{\infty}\frac{1-q^{r-\frac{1}{2}m+1}u^{-1}}{1-q^{r-\frac{1}{2}m}u}=\sum_{e\in\mathbb{Z}}\mathcal{I}_{\Delta}^{c}(m,e)u^{e}\,,\\
        &\text{where }\mathcal{I}_{\Delta}^{c}(m,e)=\sum_{n=\lfloor e\rfloor}^{\infty}\frac{(-1)^{n}q^{\frac{1}{2}n(n+1)-(n+\frac{1}{2}e)m}}{(q)_{n}(q)_{n+e}}\,.
    \end{aligned}
\end{equation}
$\mathcal{I}_{\Delta}(m,u)$ computes the generalized superconformal index of the $\mathcal{T}_{\Delta}$ theory. $(m,u)$ are (background monopole flux, fugacity) for the $U(1)$ flavor symmetry. Using the $T[SU(2)]$ index, the superconformal index of the $D(\vec{k})$ theory can be computed as follows:
\begin{equation}
\label{eq:dkindex}
    \begin{aligned}
        &\mathcal{I}^{\text{sci}}_{D(\vec{k})}(\eta,\nu)=\sum_{m_{1},\dots,m_{\sharp}\in\mathbb{Z}_{\geq 0}}\oint\prod_{I=1}^{\sharp}\left(\frac{da_{I}}{2\pi ia_{I}}\Delta(m_{I},a_{I})(a_{I}(-1)^{m_{I}})^{2k^{(I)}m_{I}}\right)\\
        &\qquad\qquad\qquad\times\left(\prod_{I=1}^{\sharp-1}\mathcal{I}_{T[SU(2)]}^{\text{sci}}(a_{I},a_{I+1},\eta,\nu;m_{I},m_{I+1})\right).
    \end{aligned}
\end{equation}
Here $\Delta(m,u)$ is the contribution from a $SU(2)$ vector multiplet:
\begin{equation}
    \begin{aligned}
        &\Delta(m,u)=\frac{1}{\text{Sym}(m)}(q^{m/2}u-q^{-m/2}u^{-1})(q^{m/2}u^{-1}-q^{-m/2}u)\,,\\
        &\text{with Sym}(m)\coloneq\begin{cases}
            2,& m=0\\
            1,& m\neq 0
        \end{cases}\,.
    \end{aligned}
\end{equation}
Also, in section \ref{sec:bg24r} we discussed gauging of the $D(\vec{k}=(2,r,-2))$ theory with $\mathbb{Z}_{2}^{\text{diag}}\subset\mathbb{Z}_{2}^{(1)}\times\mathbb{Z}_{2}^{(3)}$ 1-form symmetry. To compute the superconformal index of the gauged theory using \eqref{eq:dkindex}, we should set the monopole charge summation range as follows:
\begin{equation}
\label{eq:sumrange}
    \begin{aligned}
        &2m_{1},m_{2},2m_{3}\in\mathbb{Z}_{\geq 0}\,,\\
        &n_{1},n_{2},m_{1}+n_{1},m_{1}-m_{3}\in\mathbb{Z}\,,
    \end{aligned}
\end{equation}
where $n_{I}$ is a summation parameter for the $I$-th $\mathcal{I}_{T[SU(2)]}^{\text{sci}}$(cf. \eqref{eq:tsu2index}). \eqref{eq:sumrange} reflects the Dirac quantization condition.
\paragraph{Squashed 3-sphere partition function  \cite{Kapustin:2009kz,Jafferis:2010un,Hama:2010av}}
Squashed three-sphere geometry was introduced in \eqref{eq:squashed}. Denoting the squashing parameter as $b$, $\mathcal{Z}^{S_{b}^{3}}$ of $T[SU(2)]$ can be written as follows:
\begin{equation}
\label{eq:tsusquashed}
    \begin{aligned}
        &\mathcal{Z}^{S_{b}^{3}}_{T[SU(2)]}(M_{1},M_{2};M,\nu)=\int\frac{dZ}{\sqrt{2\pi\hbar}}\mathcal{I}_{T[SU(2)]}^{\hbar}(Z,M_{1},M_{2};W),\text{ where}\\
        &\mathcal{I}_{T[SU(2)]}^{\hbar}\coloneq\exp\left(\frac{Z^{2}+M_{1}^{2}+2M_{2}Z}{\hbar}\right)\psi_{\hbar}\left(-W+(i\pi+\frac{\hbar}{2})\right)\\
        &\qquad\qquad\quad\times\prod_{\epsilon_{1},\epsilon_{2}\in\{\pm 1\}}\psi_{\hbar}\left(\epsilon_{1}Z+\epsilon_{2}M_{1}+\frac{W}{2}+(\frac{i\pi}{2}+\frac{\hbar}{4})\right)\bigg|_{W\coloneq M+(i\pi+\hbar/2)\nu}\,,
    \end{aligned}
\end{equation}
where we defined $\hbar\coloneq 2\pi ib^{2}$. $M_{1},\,M_{2}$ are the rescaled real mass($b\times$real mass) for the $U(1)$ Cartans of $SU(2)_{L},\,SU(2)_{R}$. $M$ and $\nu$ are rescaled real mass of $U(1)_{A}$ and R-symmetry mixing parameter of $U(1)_{A}$, respectively. $\psi_{\hbar}(Z)$(called quantum dilogarithm) is $\mathcal{Z}^{S_{b}^{3}}$ of the $\mathcal{T}_{\Delta}$ theory. Note that our partition functions have overall phase ambiguity, because of the decoupled invertible TQFT, background CS levels for R-symmetry and flavor symmetries, etc. From \eqref{eq:tsusquashed}, we can write the $\mathcal{Z}^{S_{b}^{3}}$ for $D(\vec{k})$:
\begin{equation}
\label{eq:dksquashedptn}
    \begin{aligned}
        &\mathcal{Z}^{S_{b}^{3}}_{D(\vec{k})}(M,\nu)\\
        &=\int\left(\prod_{I=1}^{\sharp}\frac{\Delta(M_{I})dM_{I}}{\sqrt{2\pi\hbar}}\exp\left(\frac{k^{(I)}M_{I}^{2}}{\hbar}\right)\right)\left(\prod_{I=1}^{\sharp-1}\mathcal{Z}^{S_{b}^{3}}_{T[SU(2)]}(M_{I},M_{I+1};M,\nu)\right)\,.
    \end{aligned}
\end{equation}
$\Delta(M)$ is a contribution of the $SU(2)$ vector multiplet:
\begin{equation}
    \Delta(M)=2\sinh(M)\sinh(2\pi iM/\hbar)\,.
\end{equation}
When $b=1$ and $\nu=0$, $\mathcal{Z}^{S_{b=1}^{3}}_{T[SU(2)]}$ simplifies to \cite{Gang:2021hrd,Nishioka:2011}:
\begin{equation}
\label{eq:tsusphere}
    \mathcal{Z}^{S_{b=1}^{3}}_{T[SU(2)]}(M_{1},M_{2};M=0,\nu=0)\simeq\frac{1}{2}\frac{\sin\left(\frac{M_{1}M_{2}}{\pi}\right)}{\sinh(M_{1})\sinh(M_{2})}\,.
\end{equation}
Here $\simeq$ reflects the phase ambiguity. Using \eqref{eq:tsusphere}, we can compute the three-sphere partition function of $D(\vec{k})$:
\begin{equation}
\label{eq:dkconptn}
    \begin{aligned}
        &\mathcal{Z}^{S_{b=1}^{3}}_{D(\vec{k})}(M=0,\nu=0)\\
        &\simeq 2\int\left(\prod_{I=1}^{\sharp}\frac{dM_{I}}{2\pi}\exp\left(\frac{k^{(I)}M_{I}^{2}}{2\pi i}\right)\right)\sinh(M_{1})\sinh(M_{\sharp})\prod_{I=1}^{\sharp-1}\sin\left(\frac{M_{I}M_{I+1}}{\pi}\right)\\
        &\simeq\frac{1}{\sqrt{2^{\sharp-2}|p|}}\sin\left(\frac{\pi}{|p|}\right)\,,
    \end{aligned}
\end{equation}
where we utilized the fact that
\begin{equation}
    \begin{aligned}
        &\det\Bar{\mathcal{K}}(\vec{k})=|p|\,,\\
        &\text{where }\Bar{\mathcal{K}}_{IJ}\coloneq\begin{cases}
            1,&|I-J|=1\\
            k^{(I)},&I=J\\
            0,&\text{otherwise}
        \end{cases}\qquad(I,J=1,\dots,\sharp)\,.
    \end{aligned}
\end{equation}
Also, in  \cite{Gang:2024} it was shown that, for decoupled TQFT,
\begin{equation}
    |(\mathcal{Z}^{\text{con}}\text{ of }\text{TFT}(\vec{k}))|=\begin{cases}
        2^{-\sharp/2}\,,& p\in 2\mathbb{Z}+1\\
        2^{-(\sharp-1)/2}\,,& p\in 2\mathbb{Z}
    \end{cases}\,.
\end{equation}
Combined with \eqref{eq:dkconptn}, we obtain
\begin{equation}
\label{eq:caldconptn}
    \begin{aligned}
        |(\mathcal{Z}^{\text{con}}\text{ of }\mathcal{D}(p,q))|=\frac{|(\mathcal{Z}^{\text{con}}\text{ of }D(\vec{k}))|}{|(\mathcal{Z}^{\text{con}}\text{ of }\text{TFT}(\vec{k}))|}=\begin{cases}
            \frac{2}{\sqrt{|p|}}\sin\left(\frac{\pi}{|p|}\right)\,,& p\in 2\mathbb{Z}+1\\
            \sqrt{\frac{2}{|p|}}\sin\left(\frac{\pi}{|p|}\right)\,,& p\in 2\mathbb{Z}
        \end{cases}\,.
    \end{aligned}
\end{equation}
\paragraph{Twisted partition functions  \cite{Benini:2015noa,Benini:2016hjo,Closset:2016arn,Closset:2018ghr}}
In this paragraph we will compute the twisted partition function $\mathcal{Z}^{\mathcal{M}_{g,p}}$ \eqref{eq:ptn} of $D(\vec{k})$ theory. We start by considering the integrand of $\mathcal{Z}^{S_{b}^{3}}$ \eqref{eq:dksquashedptn} in an asymptotic limit $\hbar\rightarrow 0$  \cite{Gang:2019jut}:
\begin{equation}
    \begin{aligned}
        &\log\mathcal{I}^{\hbar}_{D(\vec{k})}\xrightarrow{\hbar\rightarrow 0}\frac{1}{\hbar}\mathcal{W}_{0}^{(\vec{k})}+\mathcal{W}_{1}^{(\vec{k})}+\dots\,,\\
        &\mathcal{W}_{0}^{(\vec{k})}(\vec{Z},\vec{M};M,\nu)=\left(\sum_{I=1}^{\sharp}(\pm 2\pi iM_{I}+k^{(I)}M_{I}^{2})\right)+\sum_{I=1}^{\sharp-1}\mathcal{W}_{0}^{T[SU(2)]}(Z_{I},M_{I},M_{I+1};M,\nu)\,,\\
        &\text{where }\mathcal{W}_{0}^{T[SU(2)]}(Z,M_{1},M_{2};M,\nu)=Z^{2}+M_{1}^{2}+2M_{2}Z+\text{Li}_{2}(e^{M+i\pi\nu})\\
        &\qquad\qquad\qquad\qquad\qquad\qquad\qquad\qquad+\sum_{\epsilon_{1},\epsilon_{2}\in\{\pm 1\}}\text{Li}_{2}(-e^{-\epsilon_{1}Z-\epsilon_{2}M_{1}-\frac{(M+i\pi (\nu-1))}{2}})\,,\\
        &\mathcal{W}_{1}^{(\vec{k})}(\vec{Z},\vec{M};M,\nu)=\sum_{I=1}^{\sharp}\log (\sinh(M_{I}))+\sum_{I=1}^{\sharp-1}\mathcal{W}_{1}^{T[SU(2)]}(Z_{I},M_{I},M_{I+1};M,\nu)\text{ with}\\
        &\mathcal{W}_{1}^{T[SU(2)]}(Z,M_{1},M_{2})=-\frac{\nu}{2}\log(1+e^{M+i\pi\nu})\\
        &\qquad\qquad\qquad\qquad\qquad+\frac{\nu-1}{4}\sum_{\epsilon_{1},\epsilon_{2}\in\{\pm 1\}}\log(1+e^{-\epsilon_{1}Z-\epsilon_{2}M_{1}-\frac{(M+i\pi (\nu-1))}{2}})\,.
    \end{aligned}
\end{equation}
We used the fact that in the limit $\hbar\rightarrow 0$,
\begin{equation}
    \log\psi_{\hbar}(Z)\xrightarrow{\hbar\rightarrow 0}\frac{\text{Li}_{2}(e^{-Z})}{\hbar}-\frac{1}{2}\log(1-e^{-Z})+\dots\,.
\end{equation}
Then the Bethe-vacua of the $D(\vec{k})$ theory can be written as
\begin{equation}
    \begin{aligned}
        &\mathcal{S}^{\text{BE}}_{D(\vec{k})}(M,\nu)=\{\vec{z},\vec{m}\,\colon\,\exp(\partial_{Z_{I}}\mathcal{W}_{0}^{(\vec{k})})\rvert_{*}=\exp(\partial_{M_{J}}\mathcal{W}_{0}^{(\vec{k})})\rvert_{*}=1,m_{J}^{2}\neq 1\}/\mathbf{W}\,,\\
        &\text{with }*\,:\,Z_{I}\rightarrow\log z_{I},\, M_{J}\rightarrow\log m_{J}\,,
    \end{aligned}
\end{equation}
where $I=1,\dots,\sharp-1$ and $J=1,\dots,\sharp$. $\mathbf{W}$ denotes a Weyl subgroup $\mathbb{Z}_{2}^{\sharp}$ of the $SU(2)^{\sharp}$ gauge symmetry
\begin{equation}
    \mathbf{W}\,:\,m_{J}\rightarrow 1/m_{J}\text{ for each }J=1,\dots,\sharp\,.
\end{equation}
Handle gluing $\mathcal{H}$ and fibering operator $\mathcal{F}$ of the $D(\vec{k})$ theory are ($\vec{X}\coloneq(\vec{Z},\vec{M})$)
\begin{equation}
\label{eq:hfcomputation}
    \begin{aligned}
        &\mathcal{H}_{D(\vec{k})}(\vec{z},\vec{m};M,\nu)=\frac{e^{i\delta}}{|\mathbf{W}|^{2}}\left(\det_{A,B}\left(\partial_{X_{A}}\partial_{X_{B}}\mathcal{W}_{0}^{(\vec{k})}\right)\right)\exp\left(-2\mathcal{W}_{1}^{(\vec{k})}\right)\bigg|_{*}\,,\\
        &\mathcal{F}_{D(\vec{k})}(\vec{z},\vec{m};M,\nu)=\exp\left(-\frac{\mathcal{W}_{0}^{(\vec{k})}-\vec{X}\cdot\partial_{\vec{X}}\mathcal{W}_{0}^{(\vec{k})}-M\partial_{M}\mathcal{W}_{0}^{(\vec{k})}}{2\pi i}\right)\bigg|_{*}\,,
    \end{aligned}
\end{equation}
where $|\mathbf{W}|=2^{\sharp}$ and $e^{i\delta}$ is a phase factor. We fix the phase ambiguity by requiring that
\begin{equation}
    \mathcal{Z}_{D(\vec{k})}^{\mathcal{M}_{g=0,p=0}}(M=0,\nu=\pm 1)=\sum_{(\vec{z},\vec{m})\in\mathcal{S}_{\text{BE}}}\mathcal{H}_{D(\vec{k})}^{-1}=1\,.
\end{equation}
For $D(\vec{k})$ theory with $\vec{k}=(k_{1},k_{2})$, there are $2(|k_{1}k_{2}-1|-1)$ Bethe-vacua whose handle gluing operators are  \cite{Gang:2021hrd}
\begin{equation}
    \{\mathcal{H}_{D(\vec{k}=(k_{1},k_{2}))}(\vec{z},\vec{m})\,:\,(\vec{z},\vec{m})\in\mathcal{S}_{\text{BE}}(M=0,\nu=\pm 1)\}=\left\{\frac{|k_{1}k_{2}-1|}{\sin^{2}\left(\frac{\pi n}{k_{1}k_{2}-1}\right)}^{\otimes 2}\right\}^{|k_{1}k_{2}-1|-1}_{n=1}\,.
\end{equation}
Also, $\textbf{HF}_{\text{bos}}\coloneq \{(\mathcal{H}^{-1/2},\mathcal{F}/\mathcal{F}_{\alpha =0})\,:\,(\vec{z},\vec{m})\in\mathcal{S}_{\text{BE}}\}$ of $D(\vec{k}=(k_{1},k_{2}))$ can be factorized as follows:
\begin{equation}
    \textbf{HF}_{\text{bos}}=\textbf{HF}'_{\text{bos}}\times\begin{cases}
        \{(\frac{1}{2},1),(\frac{1}{2},1),(\frac{1}{2},1),(\frac{1}{2},-1)\},& k_{1}\text{ and }k_{2}\text{ are both even,}\\
        \{(\frac{1}{2},1),(\frac{1}{2},1),(\frac{1}{2},i),(\frac{1}{2},i^{-1})\},&\text{one of them is odd,}\\
        \{(\frac{1}{\sqrt{2}},1),(\frac{1}{\sqrt{2}},i)\}\text{ or }\{(\frac{1}{\sqrt{2}},1),(\frac{1}{\sqrt{2}},i^{-1})\},&\text{both are odd.}
    \end{cases}
\end{equation}
For more details about the decoupled TQFT TFT[$\vec{k}$] see the previous paper  \cite{Gang:2024}. 

\section{$S^{3}_{b=1}$ partition function of $D(\Vec{k})$ with $N=3$}
According to  \cite{Gang:2015asp}, the superconformal $S^{3}_{b=1}$ partition function of $D(\Vec{k})$ can be written as 
\begin{equation}
\label{eq:s3partition}
    \begin{aligned} 
    &\mathcal{Z}^{S^{3}_{b=1}}_{D(\Vec{k}),N}(\nu=0)\\
    &=\int\left(\frac{1}{N!}\prod_{I=1}^{\sharp}d\Vec{\mu}^{I}\Delta(\Vec{\mu}^{I})^{2}e^{k^{(I)}\pi i(\Vec{\mu}^{I})^2}\delta(\sum_{i=1}^{N}\Vec{\mu}^{I}_{i})\right)\left(\prod_{I=1}^{\sharp -1}\sum_{\sigma\in G_{N}}(-1)^{\sigma}\frac{e^{2\pi i\Vec{\mu}^{I+1}\cdot\sigma(\Vec{\mu}^{I})}}{\Delta(\Vec{\mu}^{I})\Delta(\Vec{\mu}^{I+1})}\right)\\
    &=\int\left(\frac{1}{N!}\prod_{I=1}^{\sharp}d\Vec{\mu}^{I}e^{k^{(I)}\pi i(\Vec{\mu}^{I})^2}\delta(\sum_{i=1}^{N}\Vec{\mu}^{I}_{i})\right)\Delta(\Vec{\mu}^{1})\Delta(\Vec{\mu}^{\sharp})\left(\prod_{I=1}^{\sharp -1}\sum_{\sigma\in G_{N}}(-1)^{\sigma}e^{2\pi i\Vec{\mu}^{I+1}\cdot\sigma(\Vec{\mu}^{I})}\right)\,,
    \end{aligned}
\end{equation}
where $\Delta(\Vec{\mu})$ is
\begin{equation}
    \Delta(\Vec{\mu})=\prod_{i<j}2\sinh{\pi(\mu_{i}-\mu_{j})}\,.
\end{equation}
For $N=3$,
\begin{equation}
\label{eq:part}
    \begin{aligned}
        &\mathcal{Z}^{S^{3}_{b=1}}_{D(\Vec{k}),N}(\nu=0)\\
    &=\frac{1}{6}\int\left(\prod_{I=1}^{\sharp}d\Vec{\mu}^{I}e^{k^{(I)}\pi i(\Vec{\mu}^{I})^2}\delta(\sum_{i=1}^{3}\mu^{I}_{i})\right)\Delta(\Vec{\mu}^{1})\Delta(\Vec{\mu}^{\sharp})\left(\prod_{I=1}^{\sharp -1}\sum_{\sigma\in G_{3}}(-1)^{\sigma}e^{2\pi i\Vec{\mu}^{I+1}\cdot\sigma(\Vec{\mu}^{I})}\right)\\
    &=\frac{6^{\sharp-2}}{(2\pi)^{\sharp}}\int\left(\prod_{I=1}^{\sharp}d\Vec{\mu}^{I}e^{k^{(I)}\pi i(\Vec{\mu}^{I})^2}\int_{-\infty}^{\infty}dy^{I}e^{iy^{I}(\mu^{I}_{1}+\mu^{I}_{2}+\mu^{I}_{3})}\right)\left(e^{\pi(\mu^{1}_{1}-\mu^{1}_{2})}-e^{-\pi(\mu^{1}_{1}-\mu^{1}_{2})}\right)\\
    &\times\left(e^{\pi(\mu^{1}_{2}-\mu^{1}_{3})}-e^{-\pi(\mu^{1}_{2}-\mu^{1}_{3})}\right)\left(e^{\pi(\mu^{1}_{1}-\mu^{1}_{3})}-e^{-\pi(\mu^{1}_{1}-\mu^{1}_{3})}\right)\left(e^{\pi(\mu^{\sharp}_{1}-\mu^{\sharp}_{2})}-e^{-\pi(\mu^{\sharp}_{1}-\mu^{\sharp}_{2})}\right)\\
    &\times\left(e^{\pi(\mu^{\sharp}_{2}-\mu^{\sharp}_{3})}-e^{-\pi(\mu^{\sharp}_{2}-\mu^{\sharp}_{3})}\right)\left(e^{\pi(\mu^{\sharp}_{1}-\mu^{\sharp}_{3})}-e^{-\pi(\mu^{\sharp}_{1}-\mu^{\sharp}_{3})}\right)\left(\prod_{I=1}^{\sharp -1}e^{2\pi i\Vec{\mu}^{I+1}\cdot\Vec{\mu}^{I}}\right)
    \end{aligned}
\end{equation}
Let us evaluate the integral in the last line of \eqref{eq:part}. We will use the formula
\begin{equation}
    \int d^{n}x\exp\left(-\frac{1}{2}x^{T}Ax+j^{T}x\right)=\exp\left(\frac{1}{2}j^{T}A^{-1}j\right)\frac{(2\pi)^{n/2}}{\sqrt{\det A}}\,.
\end{equation}
Second-order terms in the exponent can be written as follows:
\begin{equation}
\label{eq:quadratic}
    \begin{aligned}
        &\sum_{I=1}^{\sharp}k^{(I)}\pi i(\Vec{\mu}^{I})^2+2\pi i\sum_{I=1}^{\sharp -1}\Vec{\mu}^{I+1}\cdot\Vec{\mu}^{I}=-\frac{1}{2}\Vec{\mu}^{T}\cdot A\cdot\Vec{\mu}\,,\\
        &\text{where }A_{ij}=(-2\pi i)\times\begin{cases}
        1, & |i-j|=3\\
        k^{([\frac{i+2}{3}])}, & i=j\\
        0, & \text{otherwise}
        \end{cases}\qquad(i,j=1,\cdots,3\sharp)\,,\\
        &\det A=(-2\pi i)^{3\sharp}p^{3}\,.
    \end{aligned}
\end{equation}
First-order terms in the exponent involving $y$ can be written as
\begin{equation}
\label{eq:firstorder}
    \begin{aligned}
        i\begin{pmatrix}
            y^{1}&y^{1}&y^{1}&\cdots&y^{\sharp}&y^{\sharp}&y^{\sharp}
        \end{pmatrix}\cdot\begin{pmatrix}
            \mu_{1}^{1}&\mu_{2}^{1}&\mu_{3}^{1}&\cdots&\mu_{1}^{\sharp}&\mu_{2}^{\sharp}&\mu_{3}^{\sharp}
        \end{pmatrix}^{T}\,.
    \end{aligned}
\end{equation}
When integrated over $\mu_{i}^{I}$, \eqref{eq:quadratic} and \eqref{eq:firstorder} give
\begin{equation}
\label{eq:firstorder2}
\begin{aligned}
    &\frac{(2\pi)^{3\sharp/2}}{\sqrt{\det A}}\exp\left(\frac{1}{2}j^{T}A^{-1}j\right)\\
    &=\frac{(2\pi)^{3\sharp/2}}{\sqrt{\det A}}\exp\left(-\frac{1}{2}\begin{pmatrix}
            y^{1}&y^{1}&y^{1}&\cdots&y^{\sharp}&y^{\sharp}&y^{\sharp}
        \end{pmatrix}\cdot A^{-1}\cdot\begin{pmatrix}
            y^{1}&y^{1}&y^{1}&\cdots&y^{\sharp}&y^{\sharp}&y^{\sharp}
        \end{pmatrix}^{T}\right)\,.
        \end{aligned}
\end{equation}
Integrating over $y^{I}$, the entire integral becomes(up to overall phase factor)
\begin{equation}
\label{eq:delta}
    \frac{(2\pi)^{3\sharp/2}}{\sqrt{\det A}}\times(2\pi)^{\sharp}\cdot3^{-\sharp/2}\sqrt{p}\,.
\end{equation}
First-order terms from $\sinh$ factors give(up to overall phase factor), after integrating over $\mu_{i}^{I}$,
\begin{equation}
\label{eq:sinh}
    96\sin^{3}\frac{\pi}{p}\cos\frac{\pi}{p}\,.
\end{equation}
Using \eqref{eq:delta} and \eqref{eq:sinh}, \eqref{eq:part} becomes
\begin{equation}
\label{eq:n3zcon}
    |\mathcal{Z}^{S^{3}_{b=1}}_{D(\Vec{k}),N=3}(\nu=0)|=(\text{factor dependent on }\sharp)\cdot\frac{1}{|p|}\sin^{3}\left(\frac{\pi}{|p|}\right)\cos\left(\frac{\pi}{|p|}\right)\,.
\end{equation}
Note that we didn't carefully track the overall factor dependent on $\sharp$, which may be related to the decoupled TQFT.


\bibliographystyle{ytphys}
\bibliography{ref}
	
\end{document}